\documentclass[acmsmall]{acmart}
\AtBeginDocument{%
  }

\usepackage{subcaption}
\usepackage{multirow}
\usepackage{tabularx}
\usepackage{xcolor}
\usepackage{colortbl}
\usepackage{makecell}
\usepackage{colortbl}
\usepackage{enumitem}
\usepackage{fontawesome}
\definecolor{lightgray}{gray}{0.9}
\definecolor{lightpink}{rgb}{1.0, 0.9, 0.9}
\definecolor{lightyellow}{rgb}{1.0, 1.0, 0.9}
\definecolor{lightgreen}{rgb}{0.9, 1.0, 0.9}
\definecolor{lightblue}{rgb}{0.9, 0.95, 1.0}

\newcommand{\revise}[1]{\textcolor{black}{#1}}

\setcopyright{cc}
\setcctype{by}
\acmDOI{10.1145/3832130}
\acmYear{2026}
\acmJournal{PACMSE}
\acmVolume{3}
\acmNumber{ISSTA}
\acmArticle{ISSTA039}
\acmMonth{10}
\acmSubmissionID{issta26main-p348-p}
\received{2026-01-30}
\received[accepted]{2026-04-16}

\begin{document}

\title[Lightweight Model Editing for LLMs to Correct Deprecated API Recommendations]{Don’t Use a Cannon to Kill a Fly: Lightweight Model Editing for LLMs to Correct Deprecated API Recommendations}

\author{Guancheng Lin}
\orcid{0009-0009-8339-313X}
\affiliation{%
  \institution{Department of Computer Science, City University of Hong Kong}
  \city{Hong Kong}
  \country{China}
}
\email{guanchlin4-c@my.cityu.edu.hk}

\author{Xiao Yu}
\orcid{0000-0002-4473-3068}
\authornote{Corresponding Author: Xiao Yu.}
\affiliation{
  \institution{The State Key Laboratory of Blockchain and Data Security, Zhejiang University}
  \city{Hangzhou}
  \country{China}
}
\affiliation{%
  \institution{\\Hangzhou High-Tech Zone (Binjiang) Institute of Blockchain and Data Security}
  \city{Hangzhou}
  \country{China}
}
\email{xiao.yu@zju.edu.cn}

\author{Jacky Keung}
\orcid{0000-0002-3803-9600}
\affiliation{%
  \institution{Department of Computer Science, City University of Hong Kong}
  \city{Hong Kong}
  \country{China}
}
\email{jacky.keung@cityu.edu.hk}

\author{Xing Hu}
\orcid{0000-0003-0093-3292}
\affiliation{
  \institution{The State Key Laboratory of Blockchain and Data Security, Zhejiang University}
  \city{Hangzhou}
  \country{China}
}
\affiliation{%
  \institution{\\Hangzhou High-Tech Zone (Binjiang) Institute of Blockchain and Data Security}
  \city{Hangzhou}
  \country{China}
}
\email{xinghu@zju.edu.cn}

\author{Xin Xia}
\orcid{0000-0002-6302-3256}
\affiliation{
  \institution{The State Key Laboratory of Blockchain and Data Security, Zhejiang University}
  \city{Hangzhou}
  \country{China}
}
\affiliation{%
  \institution{\\Hangzhou High-Tech Zone (Binjiang) Institute of Blockchain and Data Security}
  \city{Hangzhou}
  \country{China}
}
\email{xin.xia@acm.org}

\author{Alex X. Liu}
\orcid{0000-0002-6916-1326}
\affiliation{%
  \institution{Midea Group}
  \city{Foshan}
  \country{China}
}
\email{alexliu@midea.com}

\renewcommand{\shortauthors}{G. Lin et al.}

\begin{abstract}

Pre-trained or fine-tuned on large code corpora, Large Language Models (LLMs) have demonstrated strong performance in code completion tasks. However, their embedded knowledge is constrained by the timeliness of training data, which often includes code using deprecated APIs. Consequently, LLMs frequently generate deprecated APIs that will no longer be supported in future versions of third-party libraries. While retraining LLMs on updated codebases could refresh their API knowledge, this approach is computationally expensive. Recently, lightweight model editing methods have emerged to efficiently correct specific knowledge in LLMs. However, it remains unclear whether these methods can effectively update deprecated API knowledge and enable edited models to generate up-to-date APIs. 
To address this gap, we conduct the first systematic study applying 10 state-of-the-art model editing techniques to update deprecated API knowledge in three LLMs: Qwen2.5-Coder, CodeGemma, and DeepSeek-Coder. We introduce \texttt{EDAPIBench}, a dedicated benchmark featuring over 70 deprecated APIs from 8 popular Python libraries, with more than 3,000 editing instances. 
Our results show that the parameter-efficient fine-tuning method AdaLoRA achieves the best performance in enabling edited models to generate correct, up-to-date APIs, but falls short in Specificity (i.e., the editing influences untargeted knowledge). To resolve this, we propose AdaLoRA-L, which defines ``Common API Layers'' (layers within the LLMs with high importance across all APIs, storing general knowledge and excluded from editing) and restricts edits exclusively to ``Specific API Layers'' (layers with high importance only for the target API, storing the API-specific knowledge). Experimental results demonstrate that AdaLoRA-L significantly improves Specificity while maintaining comparable performance across other evaluation metrics.

\end{abstract}





\maketitle

\section{Introduction}

In modern software engineering, rapid iteration and increasing system complexity necessitate the efficient reuse of existing solutions. A fundamental practice is the use of third-party libraries, which provide reusable functionality through Application Programming Interfaces (APIs) ~\cite{wang2020empirical, zhan2021research}. 
While these libraries accelerate development, they continuously evolve through refactorings~\cite{kula2018empirical}, bug fixes~\cite{hu2023empirical}, security patches~\cite{wang2020exploring}, and feature enhancements, resulting in frequent API updates where older APIs are deprecated and replaced. 
For example, in PyTorch~\cite{pytorch}, a popular deep learning library, the API \texttt{torch.svd()} has been deprecated in favor of \texttt{torch.linalg.svd()} and will be removed in future releases. Consequently, new code should avoid deprecated APIs, as they often lack compatibility with new features or data formats and are eventually removed from the library~\cite{api_lifecycle_stages}.

This dynamic nature of APIs poses significant challenges for code-assistance tools powered by Large Language Models (LLMs). Although LLMs have demonstrated strong capabilities in code understanding and generation, enabling their widespread use in tasks such as code completion~\cite{izadi2024language,wang2025rag,guo2024ft2ra}, their knowledge is constrained by the timeliness of training data. Many LLMs were trained on codebases using deprecated API versions before up-to-date iterations, leaving them with inherently outdated knowledge. Alternatively, some LLMs may have been trained on mixed datasets containing both deprecated and up-to-date API usages, but their understanding of deprecated APIs may not be fully corrected. Therefore, during code completion, LLMs often suggest deprecated API invocations, leading to suboptimal or erroneous code recommendations. 
For example, Wang et al.~\cite{wang2024llms} report that 37.4\% of GPT-3.5’s API predictions are deprecated, highlighting the severity of this issue.

\noindent \textbf{Motivations.} 
To address this problem, Wang et al.~\cite{wang2024llms} proposed two methods, among which \texttt{REPLACEAPI} achieved the best performance. \texttt{REPLACEAPI} detects deprecated APIs in the LLM’s code completion using a maintained deprecated–up-to-date API mapping list, removes the deprecated API and all subsequent tokens, appends the corresponding up-to-date API to form a new prefix, and concatenates it with the original prompt to re-invoke the LLM for code completion.
However, this design introduces several practical limitations. In deployment, \texttt{REPLACEAPI} must continuously scan all generated completions to check for deprecated APIs, incurring additional detection latency even when no deprecated–up-to-date API replacement is required. As a result, code completions cannot be streamed token-by-token as in standard LLM decoding; instead, the system must wait for the full completion and post-hoc detection before returning the output, leading to a noticeably degraded user experience in interactive settings such as IDEs. Once a deprecated API is detected, \texttt{REPLACEAPI} further incurs extra inference cost by re-invoking the LLM with a modified prefix, increasing both latency and token consumption, especially in long-context scenarios. More fundamentally, \texttt{REPLACEAPI} performs only surface-level API name substitution without internalizing the semantics of updated APIs, and therefore cannot reliably ensure the correctness of downstream code generation (e.g., valid parameter usage) when API signatures or behaviors change. 

A more straightforward solution is to retrain LLMs on the latest codebases that predominantly use up-to-date APIs, thereby overwriting outdated knowledge and enabling the model to generate up-to-date APIs in a single inference pass. However, this approach—akin to “killing a fly with a cannon”—is computationally expensive and time-consuming, rendering it impractical for frequent API updates. Recently, researchers have proposed model editing as a lightweight alternative that avoids full retraining. Model editing aims to update specific pieces of knowledge inside an LLM efficiently and effectively while leaving unrelated knowledge intact~\cite{wang2024knowledge}. This technique has been primarily studied in the Natural Language Processing (NLP) domain~\cite{hartvigsen2023aging,fang2024alphaedit,li2024pmet,meng2022locating,meng2022mass,tan2024massiveeditinglargelanguage}. For example, in factual knowledge editing, a model might be updated to reflect a change in real-world facts—such as correcting the statement ``Joe Biden is the President of USA'' to ``Donald Trump is the President''—without degrading the model’s performance on other unrelated tasks.  However, it remains unclear whether existing model editing methods can effectively update deprecated API knowledge in LLMs and enable the edited models to generate up-to-date APIs.

\noindent \textbf{Approaches.} To bridge this gap, we conduct the first systematic study on applying 10 state-of-the-art model editing techniques for updating deprecated API knowledge in three LLMs (Qwen2.5-Coder (3B), CodeGemma (2B), and DeepSeek-Coder (1.3B)). To facilitate this, we first propose a dedicated \textbf{E}diting \textbf{D}eprecated \textbf{API} \textbf{Bench}mark, \texttt{EDAPIBench}, which can be \revise{mostly} automatically constructed. We begin with 145 verified API mappings (deprecated → up-to-date) from Wang et al.~\cite{wang2024llms}, covering eight popular Python libraries such as PyTorch and TensorFlow. Using these mappings, we extract 65,596 real-world functions from GitHub that call the up-to-date APIs. For each function, we extract lines preceding the API invocation as candidate editing inputs (to prompt LLM API completion) and the invocation line as the target API (ground truth). We then filter these inputs per LLM, retaining only inputs where the original model outputs the deprecated API—ensuring each case genuinely requires editing. This process yields over 900 unique LLM-specific editing instances per model. 
Next, we extend this core data to build four evaluation subsets that comprehensively assess model editing performance across four key dimensions:
\textit{Effectiveness}: Measuring whether the edited model generates the up-to-date API for the original inputs;
\textit{Generalization}: Measuring whether it generates the up-to-date API on semantically equivalent but syntactically varied inputs;
\textit{Portability}: Measuring whether it generates the up-to-date API across different inputs (with both syntactic and semantic differences) that involve the same deprecated API in their original completions;
\textit{Specificity}: Measuring whether it preserves consistent pre-editing behavior on inputs unrelated to the editing task. Among all evaluated model editing methods, the parameter-efficient fine-tuning approach AdaLoRA achieves the highest performance in Effectiveness, Generalization, and Portability. However, it exhibits poor Specificity, primarily because its edits tend to inadvertently modify parameters that encode general knowledge, leading to unintended changes in the model’s behavior on inputs unrelated to the editing task. 

To address this limitation, we propose an improved variant, AdaLoRA-L, to enhance Specificity while preserving other strengths. Its core design involves: (1) computing gradients of editable parameters for each editing instance to quantify their relevance to the target API; (2) calculating layer importance scores (average squared gradient magnitude of layer parameters) to distinguish ``Common API Layers'' (high importance across all APIs, storing general knowledge, excluded from editing) and ``Specific API Layers'' (high importance only for the target API, storing API-specific knowledge, set as editing targets); (3) restricting edits exclusively to Specific API Layers to avoid interfering with general knowledge. Results show AdaLoRA-L significantly  improves Specificity by 79.1\%, 67.6\%, 45.5\% across the three target edited models, while maintaining AdaLoRA’s strong performance in Effectiveness, Generalization, and Portability. In addition, AdaLoRA-L outperforms \texttt{REPLACEAPI} in Effectiveness, Generalization, and parameter-correct API generation by updating the model’s internal knowledge, while \texttt{REPLACEAPI} has higher inference and token costs and is more suitable when API name correctness matters more than parameter accuracy or user experience. 

In summary, our paper makes the following contributions:

    (1) \textbf{Benchmark:} 
    We construct \texttt{EDAPIBench}—the first dedicated benchmark for evaluating deprecated API knowledge editing in LLMs, with \revise{mostly} automated construction, serving as a standardized, rigorous platform that benefits both software engineering and broader NLP communities.  
    
     (2) \textbf{Evaluation:} We evaluate 10 state-of-the-art model editing techniques on three LLMs, and find AdaLoRA excels in three key dimensions but lacks Specificity due to inadvertently modifying parameters encoding general knowledge.

     (3) \textbf{Method:} We propose AdaLoRA-L, an improved model editing approach that isolates API-specific layers via gradient-based importance scoring, addressing AdaLoRA’s Specificity limitation while retaining its strengths in Effectiveness, Generalization, and Portability.

\section{Task Definition}

Our study frames the model editing scenario as a code completion task, with the primary goal of updating deprecated API knowledge within LLMs. 
In this context, we denote the pre-editing LLM as $f$ and define an editing instance as a tuple  $M=(x,y)$, where  $x$ represents the editing input (a code snippet to be completed) and  $y$ is the editing target (the specific correct, up-to-date API that should replace the deprecated one). 
For a given input  $x$, the current API completion result produced by the original LLM  $f$ is denoted as $y_{d}$, which corresponds to the deprecated API call. The goal of model editing is to enable the edited LLM (denoted $f_{e}$) to generate the desired target $y$ for the input $x$, effectively correcting the deprecated API call $y_{d}$. 
Following existing works~\cite{li2024model,huang2024can,wang2024knowledge}, we define that an effective model edit should meet the following four key criteria.

\textbf{Effectiveness} ensures that the edited LLM $f_{e}$ correctly generates the up-to-date API $y$ for the  editing input $x$: $f_{e}(x) = y.$

\textbf{Generalization} requires that for any input $x_g$ which is a syntactic variation of the original input $x$ but semantically equivalent (i.e., both $x$ and $x_g$ have the same meaning or behavior), and for which the original LLM $f$ still completes with the deprecated API $y_{d}$ for $x_g$, the edited LLM $f_{e}$  should generate the correct, up-to-date API $y$ for $x_g$: $f_{e}\left( x_{g} \right) = y.$

\textbf{Portability} requires that for any input $x_p$, which represents a real-world input that is both semantically and syntactically different from $x$ and is completed by the original LLM $f$ using the deprecated API $y_{d}$, the edited LLM $f_{e}$ should generate the up-to-date API $y$ for $x_p$: $f_{e}\left( x_{p} \right) = y.$

\textbf{Specificity}  ensures that model editing on the original LLM $f$ does not affect other irrelevant knowledge, preserving unchanged outputs for inputs irrelevant to the editing task. Let $\mathbb{U} = \left\{ {U_{1}, U_{2},\ldots} \right\}$ denote a set of unrelated instances, where each $U = \left( {x_{u},y_{u}} \right)$ consists of a non-target input $x_u$, and its corresponding API completion output by the original LLM $f$ (i.e., $y_{u}$= $f$($x_{u}$)). Here, $x_u$ is unrelated to the editing input $x$ and involves no overlapping API calls with the target API $y$ or the deprecated API $y_{d}$. The edited LLM $f_{e}$ should satisfy: $f_{e}\left( x_{u} \right) = f\left( x_{u} \right), \forall U \in \mathbb{U}.$

\section{Study Design}

\subsection{Model Editing Techniques and Subjects}
\label{sec: EditingSubjects}

The model editing methods can be categorized into four types based on the way knowledge is updated~\cite{huang2024can, wang2024knowledge}. 
\textbf{Locate-then-edit}: This paradigm first locates where knowledge is stored and then modifies the corresponding parameters. 
\textbf{Parameter-efficient Fine-tuning}: Unlike full fine-tuning, this method focuses on fine-tuning only the parameters of specific components of the model, thereby reducing computational and memory overhead. While such methods are not specifically designed for knowledge editing, several studies~\cite{huang2024can, li2024model, zhang2024comprehensive} adopt them as baseline methods, as this efficient model updating approach aligns with the requirements of model editing.
\textbf{Memory-based method}: Instead of modifying model parameters, this approach introduces an additional memory module to store new knowledge, which is queried when relevant inputs are encountered.
\textbf{Meta Learning}: This method trains a hypernetwork to predict how parameters should shift when editing a given instance. During the editing phase, for each edited instance, the hypernetwork is utilized to calculate the parameter shift, and the predicted shifts are subsequently applied to the model's parameters to perform the edit. 
We investigate 10 state-of-the-art model editing methods across these four categories, with details summarized in Table~\ref{tab:methods_description}. The selected methods include all those studied by Li et al.~\cite{li2024model}, a recent empirical study on model editing in software engineering, and additionally include AlphaEdit, LoRA, and AdaLoRA to enhance generalizability.

We select the three popular open-source code LLMs as our editing subjects, which are widely adopted in various software engineering tasks~\cite{pan2024codev,sultana2024code,ji2025causality, sun2024ai}. \textbf{Qwen2.5-Coder (3B)} ~\cite{hui2024qwen2}:  Proposed by Alibaba and released in September 2024, it is trained on 5.5 trillion tokens collected before February 2024. \textbf{CodeGemma (2B)} ~\cite{codegemma_2024}:  Developed by Google and released in April 2024, it is trained on 1 trillion tokens from publicly available code repositories dating up to early 2024. \textbf{DeepSeek-Coder (1.3B)} ~\cite{guo2024deepseek}:  Developed by DeepSeek and released in June 2024, it is trained from scratch on 2 trillion tokens collected before February 2023. 

\begin{table}[!t]
\centering
\caption{The brief descriptions of the ten model editing methods.}
\vspace{-0.5cm}
\label{tab:methods_description}
\small
\begin{tabularx}{\textwidth}{c|X}
\hline
% Methods & \makecell[c]{Description} \\
% \midrule
\multicolumn{2}{c}{\cellcolor{lightgray}\textbf{Locate-then-edit}} \\ \hline
\textbf{ROME}~\cite{meng2022locating} & It utilizes causal tracing to identify a particular middle-layer Feed Forward Network (FFN) where knowledge is stored. ROME conceptualizes model editing as a constrained least squares problem, implementing a rank-one update to the FFN's projection weights to incorporate new knowledge. \\
\hline
\textbf{MEMIT}~\cite{meng2022mass} & This is an enhanced version of ROME, allowing for multi-layer editing. It modifies key FFNs crucial for knowledge recall by distributing intended changes across these layers as key-value memories. \\
\hline
\textbf{PMET}~\cite{li2024pmet} & It analyzes the hidden states of Multi-Head Self-Attention (MHSA) and FFN, showing that MHSA acts as a knowledge extractor encoding general patterns without needing weight updates. Thus, PMET focuses on updating FFN weights for knowledge updates. \\
\hline
\textbf{AlphaEdit}~\cite{fang2024alphaedit} & It projects parameter perturbations onto the key matrix's null space to safeguard knowledge, preserving existing information undisturbed. By prioritizing minimal update errors, it maintains hidden representation stability, preventing forgetting during editing. \\
\hline
\multicolumn{2}{c}{\cellcolor{lightgray}\textbf{Parameter-efficient Fine-tuning}} \\ \hline
\textbf{FT-L}~\cite{zhu2020modifying} & It directly fine-tunes the FFN of a single layer and minimizes the impact on irrelevant data by constraining parameter norms. \\
\hline
\textbf{LoRA}~\cite{hu2022lora} & It fine-tunes the attention modules across all layers. It first freezes the model weights and introduces low-rank matrices into the attention modules, minimizing the volume of parameter updates by optimizing only these matrices. \\
\hline
\textbf{AdaLoRA}~\cite{zhang2023adalora} & As an optimized version of LoRA, it also fine-tunes the attention modules across all layers. The difference is that it dynamically allocates more parameter update budget to critical attention layers, while assigning fewer updates to less important ones. \\
\hline
\multicolumn{2}{c}{\cellcolor{lightgray}\textbf{Memory-based}} \\ 
\hline
\textbf{GRACE}~\cite{hartvigsen2023aging} & It introduces a module with a discrete key-value codebook for storing edited knowledge and a deferral mechanism. This mechanism enables the codebook to identify modified key-value pairs, whose values then replace the hidden layer output of the original model.\\
\hline
\textbf{A-GRACE}~\cite{li2024model} & This is an improved version of GRACE. By incorporating a contrastively-trained encoder, 
%it brings the keys of edited instances and generalization data closer together while keeping them farther from those of specificity data. 
it successfully boosts GRACE’s generalization without significantly impacting other performance dimensions. \\
\hline
\multicolumn{2}{c}{\cellcolor{lightgray}\textbf{Meta Learning}} \\ \hline
\textbf{MALMEN}~\cite{tan2024massiveeditinglargelanguage} & Instead of computing gradients to fine-tune model parameters, it uses a hypernetwork to calculate parameter shifts for editing instances, applying the shifts to the model. \\
\hline
\end{tabularx}
\vspace{-0.6cm}
\end{table}

\subsection{EDAPIBench Construction}
\label{sec: benchconstruction}
\begin{figure}
    \centering
    \includegraphics[width=0.98\linewidth]{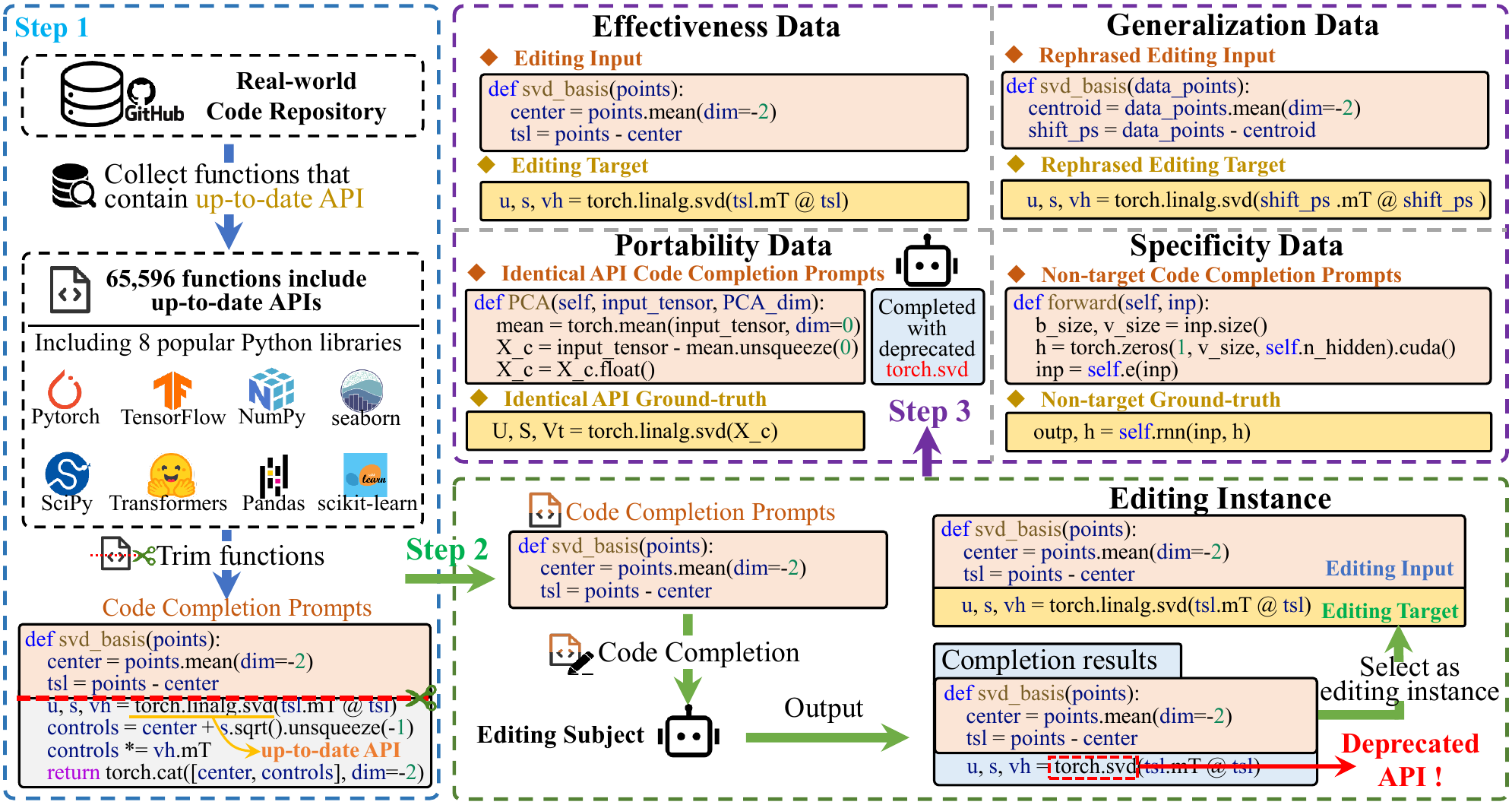}
    \captionsetup{skip=0pt}
    \caption{The construction process of \texttt{EDAPIBench}.}
    \vspace{-0.4cm}
    \label{fig:bench_construction}
\end{figure}

We construct \texttt{EDAPIBench}, which comprises over 900 editing instances $M=(x,y)$ per LLM. For each instance, the editing input $x$ is a code completion prompt—specifically, a prompt that should yield the target up-to-date API $y$, but which the original LLM $f$ completes with the deprecated API $y_{d}$. The editing target is the corresponding up-to-date API $y$. Each editing instance serves as core data to evaluate the Effectiveness of model editing; additionally, for every instance, we construct Generalization, Portability, and Specificity datasets to assess the performance of the edited LLM $f_{e}$ across the three key dimensions. Figure~\ref{fig:bench_construction} illustrates the construction process of \texttt{EDAPIBench}. 

\textbf{Step 1: Code Completion Prompts Collection.} This step focuses on collecting code snippets that can serve as editing inputs $x$—specifically, prompts that require the up-to-date API $y$ for correct completion. 
Using the 145 API mappings from Wang et al.~\cite{wang2024llms}, which pair deprecated APIs $y_d$ with up-to-date replacements $y$ across eight popular Python libraries (Figure~\ref{fig:EDAPI—libraries-num}), we search open-source repositories via the Sourcegraph tool~\cite{source-graph}, yielding 65,596 functions with calls to target up-to-date APIs. For each function, the lines preceding the API invocation serve as candidate inputs $x$, and the invocation line itself as the ground truth for the target API $y$.

\textbf{Step 2: Data Filtering.} We ensure that the three LLMs introduced in Section \ref{sec: EditingSubjects} produce the deprecated API $y_d$ for the candidate prompts before model editing. To verify this, we analyze completions by inputting each candidate editing input into each LLM separately and prompting each to complete the input three times with temperature set to 0 (using greedy sampling).  We retain only those prompts where the model outputs the deprecated API in all three completions. This yields valid, LLM-specific editing instances tailored to each model that truly require model editing. For instance, as shown in Figure~\ref{fig:bench_construction}, the editing input \texttt{def svd\_basis(points): ...tsl=points-center} is retained since the LLM generates the deprecated API \texttt{torch.svd()} in all completions.

\begin{figure}[t]
    \centering
    \hspace*{-0.75cm}
    \begin{minipage}[t]{0.55\textwidth}  
        \centering
        \includegraphics[height=2.9cm, keepaspectratio]{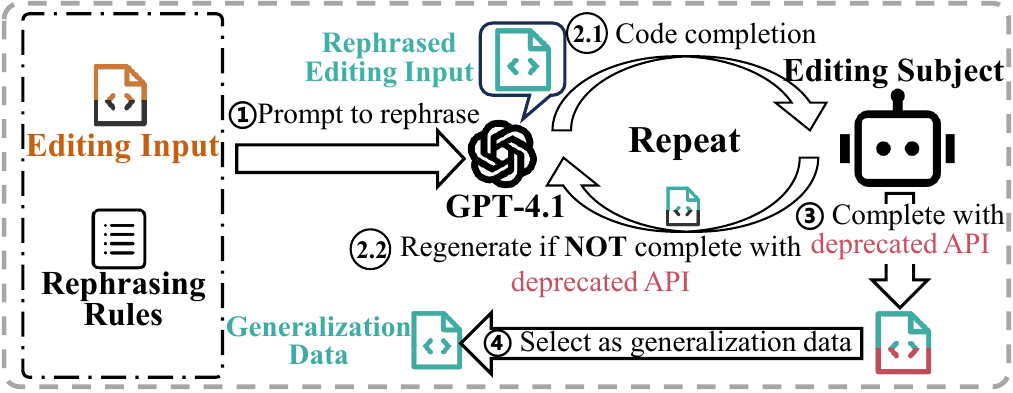} 
        \captionsetup{skip=0pt}
        \caption*{(a) Generalization Data}
    \end{minipage}%
    \begin{minipage}[t]{0.395\textwidth}  
        \centering
        \includegraphics[height=2.9cm, keepaspectratio]{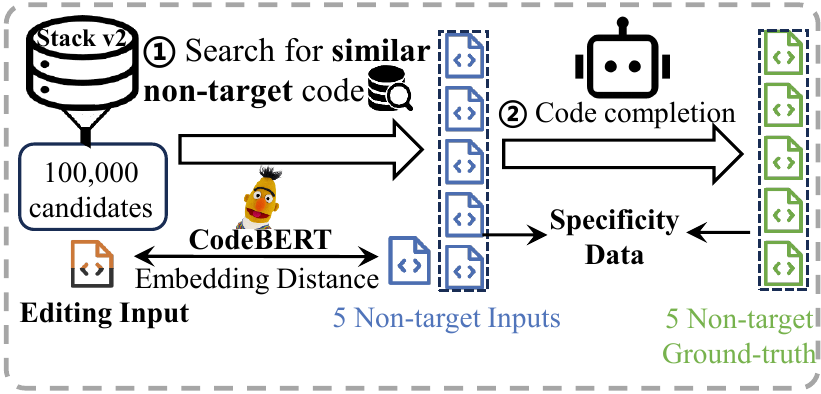} 
        \captionsetup{skip=0pt}
        \caption*{(b) Specificity Data} 
    \end{minipage}%
    \captionsetup{skip=0pt}
    \caption{The construction process of generalization data and specificity data in \texttt{EDAPIBench}.} 
    \vspace{-0.65cm}
    \label{fig:gen_spec_construction}
\end{figure}

\textbf{Step 3: Evaluation Data Preparation.}

\textit{1) Effectiveness Data.} 
The editing instances $M=(x,y)$ themselves constitute the Effectiveness data. After model editing, Effectiveness is evaluated by checking whether the edited LLM $f_{e}$ can correctly complete the prompt $x$ with the target API $y$.

\textit{2) Generalization Data.}
The generalization dimension evaluates whether the edited model $f_{e}$ can successfully complete the up-to-date API call $y$ based on code completion prompts that are semantically identical but syntactically different from the original editing input $x$. 
This requires rephrasing the code of the original editing input while preserving their meanings and behaviors. Inspired by Yu et al.~\cite{yu2022data}, who proposed 18 rephrasing rules capable of altering code while preserving semantic and syntactic naturalness, we adopt 14 of these rules, excluding four that are inapplicable to Python language (e.g., converting ``switch-case'' to ``if-else''). 
As illustrated in Figure~\ref{fig:gen_spec_construction}, and following Wang et al.~\cite{wang2024llms}, we provide these 14 rephrasing rules to GPT-4.1~\cite{GPT-4.1}, instructing it to apply them as extensively as possible to generate rephrased editing inputs that are maximally dissimilar from the original. We then prompt the editing subjects (i.e., the LLMs) to complete these rephrased inputs. If the completion does not contain the deprecated API $y_d$, GPT-4.1 is instructed to further rephrase the input, with a maximum of ten attempts. Therefore, a small number of samples still fail to elicit completions containing the deprecated API after repeated rephrasing. For all rephrased samples, the first two authors independently review them to confirm that the rephrasings are reasonable and do not alter the semantics of the context preceding the API. These validated rephrased inputs are then used as generalization data. For example, as shown in Figure~\ref{fig:bench_construction}, the rephrased input appears as 
\texttt{def svd\_basis(data\_points): ... shift\_ps=data\_points-centroid}.

\textit{3) Portability Data.} The portability dimension evaluates whether the edited model can correctly complete the up-to-date API call on different editing instances where the original model also completes the prompt with the same deprecated API. 
For each editing instance $M=(x,y)$, we randomly select another editing instances with the same target API $y$ from EDAPIBench as portability evaluation data.  For example, as shown in Figure~\ref{fig:bench_construction}, one such selected editing instance is: $x$= ``\texttt{def PCA (self, input\_tensor, PCA\_dim): X\_c=X\_c.float()} '', $y$=``\texttt{torch.linalg.svd()}''. 

\textit{4) Specificity Data.} 
The Specificity dimension requires the edited LLM $f_{e}$ to preserve consistent outputs for non-target inputs—defined as inputs for which the original LLM $f$ does not generate either the target API $y$ or the deprecated API $y_{d}$ during completion. 
In line with Wang et al.~\cite{wang2024llms}, our benchmark focuses on the most challenging scenario: whether $f_{e}$ exhibits output shifts for non-target inputs that are semantically similar to the original editing input $x$. As shown in Figure~\ref{fig:gen_spec_construction}, we randomly select 100,000 Python files from The Stack v2 dataset~\cite{lozhkov2024starcoder} as non-target input candidates. Since The Stack v2 has an extremely large scale, full retrieval would be computationally expensive, so this sampling approach is adopted for practicality. To collect semantically similar non-target inputs, we employ CodeBERT~\cite{feng2020codebert}—a widely used code embedding model in software engineering—to generate embeddings for each editing input $x$. For every $x$, we select five non-target inputs with the smallest embedding distance from the candidates, while ensuring that these inputs contain neither the target API $y$ nor the deprecated API $y_{d}$ from the corresponding editing instance. We then instruct the original LLMs (editing subjects) to complete these five non-target inputs prior to model editing, and preserve their API outputs as ``non-target ground truth''. These non-target inputs and their corresponding ground truth together constitute the Specificity data. For example, as shown in Figure~\ref{fig:bench_construction}, the non-target API for one piece of Specificity data is \texttt{self.nn()}.

\textbf{Dataset Statistics.}  
Table~\ref{tab:bench_statistics} lists the editing instances and APIs in \texttt{EDAPIBench}. After excluding APIs that LLMs cannot reliably generate in deprecated form, 82 unique deprecated–up-to-date API pairs remain from the original 145 mappings provided by Wang et al.~\cite{wang2024llms}. Generalization data matches Effectiveness in volume, as each instance has a rephrased counterpart. Specificity data is five times larger, with five non-target inputs per instance. Portability data is slightly smaller, since some APIs have only one editing instance—preventing the selection of another instance with the same deprecated API. 
Figure \ref{fig:EDAPI—libraries-num} illustrates the distribution of editing instances across libraries. 

\begin{table}[t]
    \centering
    
    \begin{minipage}{1.0\linewidth}
        \centering
        
        % \begin{table}[!t]
% \caption{The number of instances and APIs for different LLMs across evaluation dimensions in \texttt{EDAPIBench}.}
% \vspace{-0.2cm}
% \label{tab:bench_statistics}
% \small
% \begin{tabular}{c|cc|cc|cc}
% \toprule
% Models            & \multicolumn{2}{c|}{DeepSeek-Coder} & \multicolumn{2}{c|}{Qwen2.5-Coder} & \multicolumn{2}{c}{StarCoder2} \\ \midrule
% Statistical value & \# instances         & \# APIs        & \# instances        & \# APIs        & \# instances      & \# APIs      \\ \midrule
% Effectiveness     & 1,072               & 76            & 1,401              & 75            & 1,016            & 76          \\
% Generalization    & 1,072               & 76            & 1,401              & 75            & 1,016            & 76          \\
% Portability       & 1,064               & 65            & 1,390              & 67            & 1,001            & 62          \\
% Specificity       & 5,360               & 2,626         & 7,005              & 3,375         & 5,080            & 2,672       \\\bottomrule
% \end{tabular}
% \end{table}
\caption{The number of instances and APIs for different LLMs across evaluation dimensions in \texttt{EDAPIBench}.\vspace{-0.3cm}}
\label{tab:bench_statistics}
\footnotesize
\begin{tabular}{c|cc|cc|cc}
\toprule
Models          &  \multicolumn{2}{c|}{Qwen2.5-Coder} & \multicolumn{2}{c|}{CodeGemma} & \multicolumn{2}{c}{DeepSeek-Coder} \\ \hline
Statistical value & \# instances         & \# APIs        & \# instances        & \# APIs        & \# instances      & \# APIs      \\ \hline
Effectiveness     & 1,401 & 75 & 905 & 71 & 1,072 & 76          \\
Generalization    &  1,401 & 75 & 905 & 71 & 1,072 & 76          \\
Portability       & 1,390 & 67 & 892 & 58 & 1,064 & 65          \\
Specificity       & 7,005 & 3,375 & 4,525 &2,167 & 5,360 & 2,626      \\\bottomrule
\end{tabular}

% \begin{table}[!t]
% \caption{The number of instances and APIs for different LLMs across evaluation dimensions in \texttt{EDAPIBench}.}
% \vspace{-0.2cm}
% \label{tab:bench_statistics}
% \small
% \begin{tabular}{c|cc|cc|cc}
% \toprule
% Models & \multicolumn{2}{c|}{Qwen2.5-Coder} & \multicolumn{2}{c|}{DeepSeek-Coder} & \multicolumn{2}{c}{StarCoder2} \ \midrule
% Statistical value & # instances & # APIs & # instances & # APIs & # instances & # APIs \ \midrule
% Effectiveness & 1,401 & 75 & 1,072 & 76 & 1,016 & 76 \
% Generalization & 1,401 & 75 & 1,072 & 76 & 1,016 & 76 \
% Portability & 1,390 & 67 & 1,064 & 65 & 1,001 & 62 \
% Specificity & 7,005 & 3,375 & 5,360 & 2,626 & 5,080 & 2,672 \\bottomrule
% \end{tabular}
% \end{table}
    \end{minipage}

    \begin{minipage}{1.0\linewidth}
        \centering
        
    \centering
    \begin{subfigure}[t]{0.21\textwidth}
        \centering
        \includegraphics[height=2.8cm]{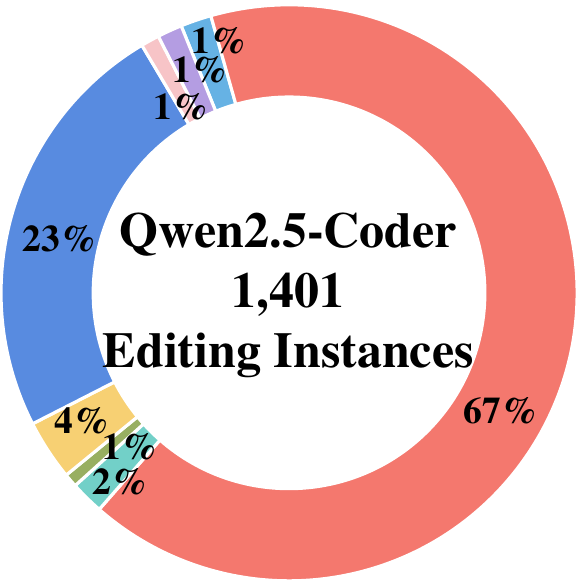}
        \label{fig:qwen—benchnum}
    \end{subfigure}
    \begin{subfigure}[t]{0.21\textwidth}
        \centering
        \includegraphics[height=2.8cm]{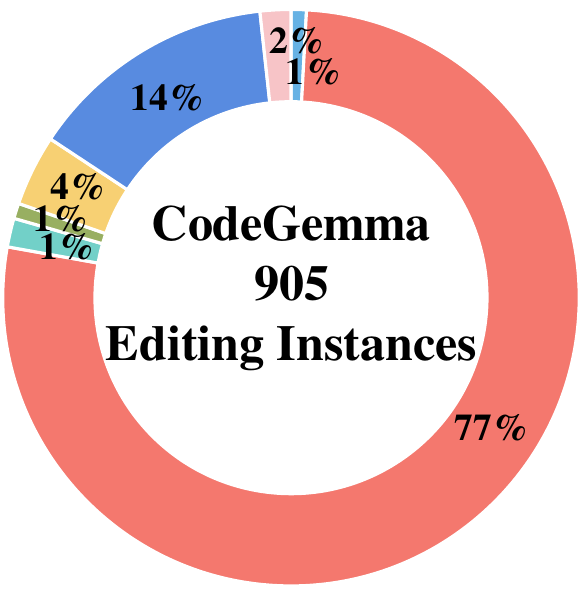}
        \label{fig:codegemma—benchnum}
    \end{subfigure}
    \begin{subfigure}[t]{0.21\textwidth}
        \centering
        \includegraphics[height=2.8cm]{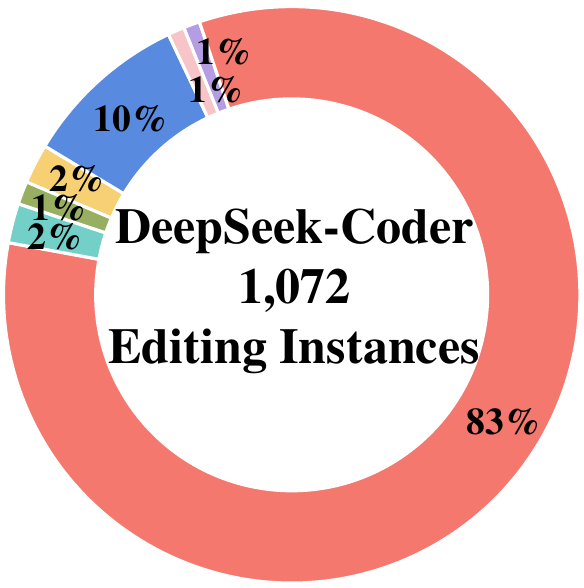}
        \label{fig:deepseek—benchnum}
    \end{subfigure}
    \begin{subfigure}[t]{0.21\textwidth}
        \centering
        \includegraphics[height=2.8cm]{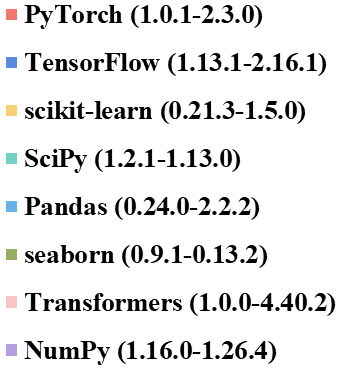}
    \end{subfigure}
    \vspace{-0.4cm}
    \captionof{figure}{The distribution of the number of target APIs from different libraries in \texttt{EDAPIBench}.}
    \captionsetup{skip=-3pt}
    \label{fig:EDAPI—libraries-num}
    \vspace{-0.6cm}

    \end{minipage}
    
\end{table}

\subsection{Experiment Settings}

All experiments are conducted on RTX 4090 GPUs paired with an Intel Core i9 processor. Following existing code completion works~\cite{yu2024fight, zhang2023repocoder}, we use incomplete code snippets directly as prompts for code completion, without adding any additional instructions. 
For the method requiring training (i.e., MALMEN), we follow existing works~\cite{li2024model,zhong2023mquake,cohen2024evaluating}, splitting EDAPIBench into training and test sets at a 1:1 ratio using a two-fold cross-validation approach, and the results of the two rounds of testing are integrated to serve as the final results.
Regarding the selection of editing layers for locate-then-edit methods, we follow Li et al.~\cite{li2024model} and maintain consistency with prior benchmarks~\cite{wang2024knowledge,wang2023easyedit}. Since these benchmarks have already identified the key editing layers via causal intervention~\cite{meng2022locating}, we directly edit the same layers and therefore do not perform additional causal intervention in our experiments.
For the hyperparameters of model editing methods, we keep them consistent with the settings used in existing model editing benchmarks~\cite{li2024model,wang2024knowledge,wang2023easyedit}. 
Our preliminary experiments indicate that varying the hyperparameters and editing layers for locate-then-edit methods does not lead to significant performance changes, and the overall conclusions remain unchanged, with AdaLoRA consistently outperforming other baselines across the first three dimensions.

\textbf{ROME, MEMIT, PMET, AlphaEdit:}
In the case of ROME, we set the 5th layer as the editing layer, while for the other three methods, we choose the \{4, 5, 6, 7, 8\}-th layers as the editing layers. All methods are trained with 25 iterations and a learning rate of 5e-1.

\textbf{GRACE, A-GRACE:}
We insert an adapter into the down projection matrix of an MLP from the later layers. The initial deferral radius of the key is set to 1, and the value corresponding to this key is fine-tuned for 30 iterations with a learning rate of 1. For the additional encoder introduced in A-GRACE, we follow the settings in the original study ~\cite{li2024model}, with a hidden layer dimension of 256, and train it for 100 epochs at a learning rate of 1e-4.

\textbf{MALMEN:}
The projection matrices of the last 5 layers of LLMs serve as the update targets for MALMEN. Initially, a hypernetwork containing two MLPs with a hidden layer dimension of 1920 is trained on the training set for predicting parameter shifts. This training process involves 1,000 iterations and a learning rate of 1e-5. During editing, the hidden states of the editing instance at the editing layers are cached first. Subsequently, the hypernetwork predicts the parameter shifts to facilitate the weight update process.

\textbf{FT-L, LoRA, AdaLoRA:}
We set an MLP from the later layers as the editing layer for FT-L to conduct fine-tuning, with 40 epochs and a learning rate of 5e-4. For LoRA and AdaLoRA, we set 30 training epochs, a rank of 8, and a learning rate of 5e-3.

The performance of model editing is evaluated based on how closely the single-line completion outputs of edited LLMs match the ground-truth.  Following Li et al.~\cite{li2024model}, we utilize three metrics: \textbf{Exact Match} (EM), \textbf{BLEU}~\cite{papineni2002bleu}, and \textbf{ROUGE-L}~\cite{lin2004rouge}. Additionally, we introduce \textbf{API Exact Match} (AEM), specifically tailored for the EDAPIBench scenario. EM measures whether the LLM’s single-line completion exactly matches the entire ground-truth line at the token level, assigning a score of 1 for a perfect match and 0 otherwise. AEM is a variant of EM that only checks whether the single-line completion contains the ground-truth API, regardless of other tokens. BLEU~\cite{papineni2002bleu} and ROUGE-L~\cite{lin2004rouge} are two commonly adopted metrics in code-related tasks~\cite{wan2018improving, ma2023attsum, bansal2021project, zhang2022improving, zhang2023diverse}. 
To reduce randomness, all methods are run five times, and the median is reported as the final performance on \texttt{EDAPIBench}.

\section{Experimental Results}

\subsection{RQ1: How do the methods perform at editing deprecated API knowledge in LLMs?}

\begin{table*}[!t]
    \centering    
    \footnotesize\tabcolsep=2.5pt
\caption{The performance of the methods on editing deprecated API knowledge within Qwen2.5-Coder.}
\label{tab: qwen}
\vspace{-0.4cm}
\begin{tabular}{c|cccc|cccc|cccc|cccc}
\hline
\multirow{2}{*}{Editor} & \multicolumn{4}{c|}{Effectiveness}                                                                  & \multicolumn{4}{c|}{Generalization}                                                                 & \multicolumn{4}{c|}{Portability}                                                                    & \multicolumn{4}{c}{Specificity}                                                                    \\ \cline{2-17}
                        & \multicolumn{1}{c}{AEM} & \multicolumn{1}{c}{EM} & \multicolumn{1}{c}{BU} & \multicolumn{1}{c|}{RL} & \multicolumn{1}{c}{AEM} & \multicolumn{1}{c}{EM} & \multicolumn{1}{c}{BU} & \multicolumn{1}{c|}{RL} & \multicolumn{1}{c}{AEM} & \multicolumn{1}{c}{EM} & \multicolumn{1}{c}{BU} & \multicolumn{1}{c|}{RL} & \multicolumn{1}{c}{AEM} & \multicolumn{1}{c}{EM} & \multicolumn{1}{c}{BU} & \multicolumn{1}{c}{RL} \\ \hline
\rowcolor{lightgray}
Pre-editd               & 2.7                     & 0.7                    & 40.4                   & 63.4                    & 5.1                     & 1.0                    & 36.9                   & 58.2                    & 2.6                     & 0.7                    & 39.8                   & 62.9                    & 99.3                    & 93.7                  & 97.6                   & 97.2                   \\ \hline
\rowcolor{lightpink}
ROME                    & 2.0                     & 0.4                    & 6.6                    & 17.5                    & 2.9                     & 0.9                    & 7.9                    & 18.8                    & 4.4                     & 1.8                    & 15.0                   & 29.1                    & 43.8                     & 24.7                    & 40.9                   & 52.2                   \\
\rowcolor{lightpink}
MEMIT                   & 6.5                     & 2.3                    & 26.4                   & 46.0                    & 6.7                     & 2.4                    & 26.1                   & 44.4                    & 4.8                     & 1.3                    & 36.3                   & 58.0                    & 91.4                    & 81.6                   & 89.1                   & 91.9                   \\
\rowcolor{lightpink}
PMET                    & 5.9                     & 2.2                    & 35.0                   & 56.3                    & 7.2                     & 2.1                    & 33.1                   & 53.0                    & 3.4                     & 0.9                    & 38.6                   & 61.4                    & 96.2                    &  88.6                 & 94.4                   & 95.1                   \\
\rowcolor{lightpink}
AlphaEdit               & 7.1                     & 2.9                    & 15.8                   & 31.9                    & 6.7                     & 2.6                    & 17.8                   & 33.5                    & 5.8                     & 1.5                    & 36.2                   & 58.3                    & 85.2                    & 72.6                   & 82.6                   & 87.2                   \\
\rowcolor{lightblue}
GRACE                   & 97.8                    & 55.5                   & 73.0                   & 80.9                    & 6.3                     & 1.4                    & 37.3                   & 58.4                    & 2.8                     & 1.0                    & 39.9                   & 62.9                    & \textbf{99.2}           & \textbf{93.4}          & \textbf{97.5}          & \textbf{97.2}   
\\
\rowcolor{lightblue}
A-GRACE                   & 96.2                    & 59.7                   & 75.6                   & 83.4                    & 88.3                     & 37.3                    & 65.2                   & 74.1                    & 47.6                    & 10.5                   & 30.0                   & 38.2                    & 97.8           & 89.6          & 97.7          & 97.6  
\\
\rowcolor{lightyellow}
MALMEN                  & 53.8                    & 15.1                   & 46.0                   & 60.7                    & 48.6                    & 10.1                   & 40.5                   & 54.1                    & 42.9                    & 9.4                    & 41.9                   & 57.7                    & 61.2                    & 41.2                    & 57.9                   & 68.4                   \\
\rowcolor{lightgreen}
FT-L                    & 58.4                    & 33.8                   & 67.4                   & 78.7                    & 48.8                    & 22.3                   & 56.2                   & 69.7                    & 16.2                    & 5.9                    & 44.8                   & \textbf{65.6}           & 96.9                    & 90.1                   & 95.0                   & 95.5                   \\
\rowcolor{lightgreen}
LoRA                    & 24.0                    & 0.0                    & 9.7                    & 26.1                    & 17.4                    & 0.0                    & 6.6                    & 20.2                    & 7.8                     & 0.0                    & 2.3                    & 10.6                    & 5.5                     & 1.2                    & 0.1                    & 1.3                    \\
\rowcolor{lightgreen}
AdaLoRA                 & \textbf{98.1}           & \textbf{60.4}          & \textbf{81.0}          & \textbf{87.6}           & \textbf{89.5}           & \textbf{39.3}          & \textbf{69.9}          & \textbf{77.7}           & \textbf{71.4}           & \textbf{11.7}          & \textbf{49.7}          & 64.2                    & 49.2                     & 28.4                    & 47.9                   & 58.3                   \\ \hline
\end{tabular}

    \footnotesize\tabcolsep=2.5pt
\caption{The performance of the methods on editing deprecated API knowledge within CodeGemma.}
\label{tab: starcoder}
\vspace{-0.4cm}
\begin{tabular}{c|cccc|cccc|cccc|cccc}
\hline
\multirow{2}{*}{Editor} & \multicolumn{4}{c|}{Effectiveness}                                                                  & \multicolumn{4}{c|}{Generalization}                                                                 & \multicolumn{4}{c|}{Portability}                                                                    & \multicolumn{4}{c}{Specificity}                                                                    \\ \cline{2-17}
                        & \multicolumn{1}{c}{AEM} & \multicolumn{1}{c}{EM} & \multicolumn{1}{c}{BU} & \multicolumn{1}{c|}{RL} & \multicolumn{1}{c}{AEM} & \multicolumn{1}{c}{EM} & \multicolumn{1}{c}{BU} & \multicolumn{1}{c|}{RL} & \multicolumn{1}{c}{AEM} & \multicolumn{1}{c}{EM} & \multicolumn{1}{c}{BU} & \multicolumn{1}{c|}{RL} & \multicolumn{1}{c}{AEM} & \multicolumn{1}{c}{EM} & \multicolumn{1}{c}{BU} & \multicolumn{1}{c}{RL} \\ \hline
\rowcolor{lightgray}
Pre-editd               & 1.0  & 0.4  & 37.6 & 60.8 & 4.5  & 1.6  & 29.4 & 49.4 & 0.8  & 0.3  & 36.2 & 59.7 & 99.5 & 98.3 & 99.2 & 96.8                   \\ \hline
\rowcolor{lightpink}
ROME                    & 1.2  & 0.0  & 1.9  & 12.4 & 0.9  & 0.1  & 1.8  & 9.8  & 6.6  & 1.1  & 17.8 & 34.0 & 42.8  & 28.5  & 38.4 & 49.3                  \\
\rowcolor{lightpink}
MEMIT                   & 3.5  & 0.1  & 4.5  & 17.5 & 4.2  & 0.7  & 3.6  & 14.5 & 5.3  & 1.0  & 31.0 & 52.2 & 89.8 & 85.4 & 88.8 & 90.2                   \\
\rowcolor{lightpink}
PMET                    & 9.4  & 1.6  & 16.1 & 33.7 & 8.5  & 1.7  & 14.3 & 29.0 & 3.7  & 0.9  & 34.5 & 57.3 & 95.2 & 91.7 & 94.2 & 93.8                   \\
\rowcolor{lightpink}
AlphaEdit               & 4.2  & 0.1  & 4.7  & 18.0 & 3.2  & 0.4  & 4.0  & 14.5 & 5.4  & 1.2  & 32.7 & 54.1 & 88.4 & 83.6 & 87.1 & 89.1                   \\
\rowcolor{lightblue}
GRACE                   & 24.2 & 12.3 & 46.7 & 65.3 & 4.6  & 1.6  & 29.3 & 49.4 & 1.0  & 0.3  & 36.7 & 60.1 & 97.3 & \textbf{94.7} & \textbf{96.8} & 95.3          \\
\rowcolor{lightblue}
A-GRACE                   & 24.1 & 11.9 & 48.7 & 66.3 & 18.1 & 4.9  & 33.8 & 52.1 & 1.8  & 0.4  & 36.3 & 59.2 & \textbf{97.5} & 94.2 & 96.6 & \textbf{95.4}          \\
\rowcolor{lightyellow}
MALMEN                  & 34.7 & 8.2  & 32.3 & 44.7 & 25.2 & 4.0  & 22.8 & 34.7 & 37.6 & 7.7  & 38.3 & 52.4 & 49.3  & 32.4  & 45.2 & 57.2                   \\
\rowcolor{lightgreen}
FT-L                    & 89.0 & 91.3 & 95.0 & 96.3 & 69.4 & 33.7 & 59.8 & 67.7 & 52.5 & 13.2 & \textbf{52.0} & \textbf{68.0} & 88.7 & 81.4 & 87.2 & 89.2                   \\
\rowcolor{lightgreen}
LoRA                    & 3.2  & 0.0  & 2.4  & 12.6 & 2.9  & 0.0  & 1.5  & 9.4  & 2.2  & 0.0  & 1.1  & 7.3  & 8.5  & 1.1  & 0.1  & 0.5                    \\
\rowcolor{lightgreen}
AdaLoRA                 & \textbf{99.0} & \textbf{91.7} & \textbf{95.3} & \textbf{96.7} & \textbf{78.7} & \textbf{39.4} & \textbf{65.1} & \textbf{71.3} & \textbf{74.2} & \textbf{13.6} & 49.2 & 61.9 & 51.2  & 34.0  & 46.5 & 56.4                  \\ \hline
\end{tabular}
    \footnotesize\tabcolsep=2.5pt
\caption{The performance of the methods on editing deprecated API knowledge within DeepSeek-Coder.}
\label{tab: deepseek}
\vspace{-0.4cm}
\begin{tabular}{c|cccc|cccc|cccc|cccc}
\hline
\multirow{2}{*}{Editor} & \multicolumn{4}{c|}{Effectiveness}                                                                    & \multicolumn{4}{c|}{Generalization}                                                                   & \multicolumn{4}{c|}{Portability}                                                                      & \multicolumn{4}{c}{Specificity}                                                                      \\ \cline{2-17}
                        & \multicolumn{1}{c}{AEM} & \multicolumn{1}{c}{EM} & \multicolumn{1}{c}{BU} & \multicolumn{1}{c|}{RL} & \multicolumn{1}{c}{AEM} & \multicolumn{1}{c}{EM} & \multicolumn{1}{c}{BU} & \multicolumn{1}{c|}{RL} & \multicolumn{1}{c}{AEM} & \multicolumn{1}{c}{EM} & \multicolumn{1}{c}{BU} & \multicolumn{1}{c|}{RL} & \multicolumn{1}{c}{AEM} & \multicolumn{1}{c}{EM} & \multicolumn{1}{c}{BU} & \multicolumn{1}{c}{RL} \\ \hline
\rowcolor{lightgray}
Pre-editd                & 3.3                       & 0.8                    & 37.2                   & 59.2                    & 4.3                       & 0.8                    & 35.3                   & 55.5                    & 2.2                       & 0.4                    & 37.7                   & 60.0                    & 86.4                      & 78.7                   & 85.8                   & 88.5                   \\ \hline
\rowcolor{lightpink}
ROME                    & 0.3                       & 0.0                    & 0.4                    & 3.4                     & 0.2                       & 0.0                    & 0.3                    & 3.0                     & 1.3                       & 0.3                    & 5.0                    & 14.5                    & 32.4                       & 19.9                    & 28.5                   & 38.9                   \\
\rowcolor{lightpink}
MEMIT                   & 4.8                       & 0.7                    & 21.5                   & 42.3                    & 4.3                       & 0.6                    & 20.4                   & 39.3                    & 3.7                       & 0.4                    & 35.8                   & 57.7                    & 83.9                      & 75.4                   & 83.4                   & 86.9                   \\
\rowcolor{lightpink}
PMET                    & 5.5                       & 0.8                    & 32.6                   & 54.4                    & 5.9                       & 1.2                    & 31.7                   & 51.5                    & 2.6                       & 0.5                    & 37.5                   & 59.5                    & 85.4                      & 77.6                   & 84.7                   & 87.9                   \\
\rowcolor{lightpink}
AlphaEdit               & 0.8                       & 0.1                    & 1.1                    & 7.7                     & 0.4                       & 0.0                    & 1.0                    & 7.1                     & 4.0                       & 0.4                    & 27.8                   & 47.7                    & 67.2                      & 56.3                   & 65.8                   & 73.2                   \\
\rowcolor{lightblue}
GRACE                   & 71.6                      & 39.8                   & 73.0                   & 83.9                    & 6.9                       & 1.9                    & 36.7                   & 56.8                    & 2.2                       & 0.4                    & 37.8                   & 60.1                    & \textbf{86.3}             & \textbf{78.6}          & \textbf{85.8}          & \textbf{88.4}          \\
\rowcolor{lightblue}
A-GRACE                   & 73.3                      & 40.4                   & 72.8                   & 83.9                    & 63.2                       & 21.3                    & 57.7                   & 70.0                    & 43.1                       & 4.9                    & 40.7                   & 56.8                    & 83.1            & 75.2          & 85.5          & 88.2          \\
\rowcolor{lightyellow}
MALMEN                  & 38.6                      & 4.8                    & 35.0                   & 49.9                    & 37.8                      & 4.7                    & 32.0                   & 45.8                    & 38.6                      & 4.5                    & 37.2                   & 53.1                    & 56.4                       & 39.5                    & 51.1                   & 61.9                   \\
\rowcolor{lightgreen}
FT-L                    & 57.2                      & 12.7                   & 54.0                   & 69.1                    & 52.4                      & 9.1                    & 48.3                   & 63.3                    & 40.8                      & 6.7                    & \textbf{47.8}                   & \textbf{65.2}           & 86.1                      & 78.3                   & 85.5                   & 88.3                   \\
\rowcolor{lightgreen}
LoRA                    & 7.7                       & 0.9                    & 6.5                    & 12.1                    & 7.7                       & 0.4                    & 5.0                    & 9.8                     & 4.2                       & 0.0                    & 1.5                    & 4.2                     & 8.5                       & 1.4                    & 0.1                    & 0.7                    \\
\rowcolor{lightgreen}
AdaLoRA                 & \textbf{98.3}             & \textbf{89.5}          & \textbf{94.7}          & \textbf{96.5}           & \textbf{93.9}             & \textbf{56.9}          & \textbf{84.2}          & \textbf{88.1}           & \textbf{58.0}             & \textbf{7.8}           & 45.5          & 61.5                    & 55.0                       & 37.5                    & 54.6                   & 64.3                   \\ \hline
\end{tabular}
\vspace{-0.6cm}

\end{table*}

As shown in Tables \ref{tab: qwen}, \ref{tab: starcoder}, and \ref{tab: deepseek}, no single method consistently excels across all four dimensions: Effectiveness, Generalization, Portability, and Specificity. Take AdaLoRA as an example—a standout method that achieves the best overall performance in three dimensions (Effectiveness, Generalization, and Portability). However, its Specificity is noticeably lower than that of methods like GRACE. Conversely, while GRACE excels in Specificity, it exhibits a clear performance gap compared to AdaLoRA in Generalization and Portability. Although A-GRACE improves upon GRACE’s Generalization, the gap between A-GRACE and AdaLoRA in Portability remains considerable, with a difference of 0.149 to 0.724 in AEM. This limitation likely stems from its contrastive training process: the model primarily optimizes for generalization data involving simple rephrasing, where the core context of the input remains largely consistent with the original editing instance. Consequently, it only learns to adapt to such straightforward rephrasing patterns. When confronted with portability data—inputs that differ substantially from the original editing input in both semantics and syntax, A-GRACE struggles to effectively apply the edited API knowledge to these more complex, scenario-specific cases. 
Additionally, compared to AdaLoRA, LoRA was originally designed as a general-purpose parameter-efficient fine-tuning method suitable for tasks such as text classification and general generation. Its goal is low-cost adaptation rather than precise knowledge updating. Regardless of the importance of parameters to the target API knowledge, LoRA adjusts all parameters using low-rank matrices of the same dimension, which leads to suboptimal overall performance in this context.  Although locate-then-edit methods have demonstrated strong performance in natural language processing tasks, they perform poorly on our benchmark. These methods are designed to edit factual knowledge in natural language, which can often be represented as explicit “subject–relation–object” triples (e.g., \textless US President, is, Trump\textgreater). Accordingly, they treat MLP layers in LLMs as key–value memories, where the subject and relation serve as the key and the object as the value. Following Li et al.~\cite{li2024model}, we adopt API-preceding code context as the key and the target API line as the value in the code domain. However, unlike the clear semantic mappings in natural language, the relationship between diverse code contexts and APIs is far more implicit. This lack of well-defined semantic triples makes locate-then-edit methods inherently ill-suited for deprecated API updates. Meta-learning methods, on the other hand, show no clear advantages or obvious shortcomings across the four dimensions, resulting in mediocre overall performance.

\begin{center} 
\vspace{-10pt}
\resizebox{\linewidth}{!}{
\begin{tabular}{!{\vrule width 1.5pt}p{1\columnwidth}}
     \noindent
     {\large \faLightbulbO} \textbf{Answer to RQ1:} No single method performs well across all four dimensions. AdaLoRA performs best in Effectiveness, Generalization, and Portability, but underperforms in Specificity. 
\end{tabular}}
\vspace{-10pt}
\end{center}

\subsection{RQ2: What is the efficiency of the methods for deprecated API knowledge editing?}

\begin{figure}[t]
    \centering
    \begin{minipage}[t]{0.49\textwidth}
        \centering
        \includegraphics[width=0.99\linewidth]{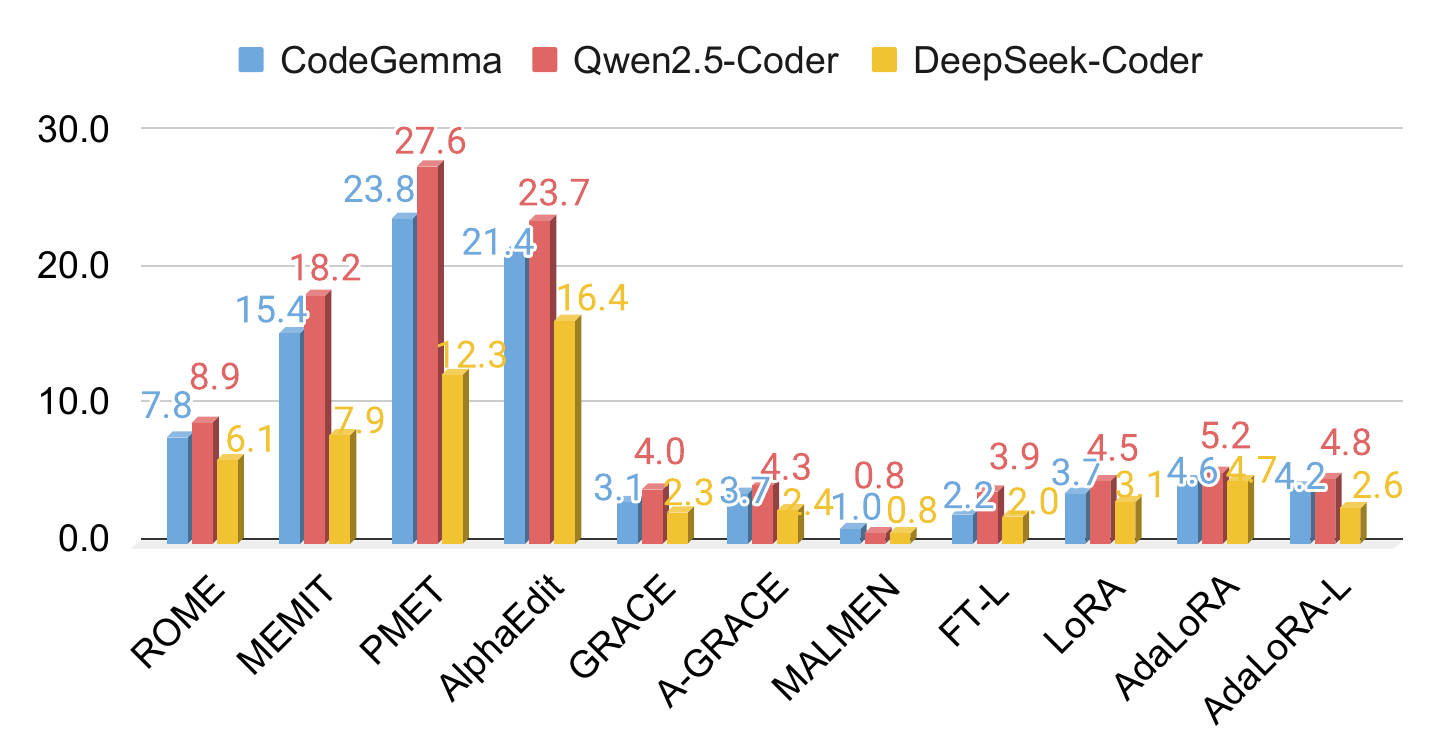}
        \captionsetup{skip=0pt}
        \caption{The average time cost (seconds) of the model editing methods per edit.}
        \label{fig:time_cost}
    \end{minipage}
    \begin{minipage}[t]{0.49\textwidth}
        \centering
        \includegraphics[width=0.99\linewidth]{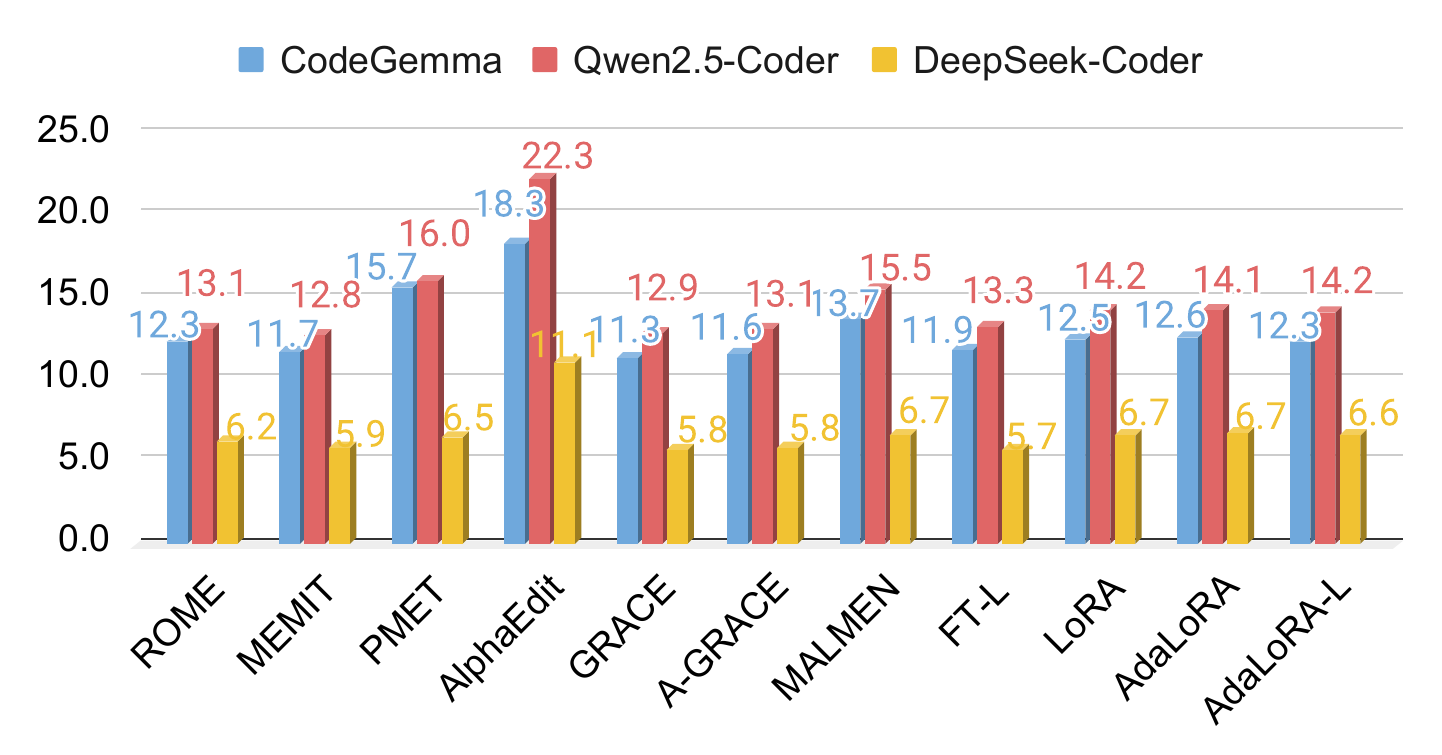}
        \captionsetup{skip=0pt}
        \caption{The average peak memory cost (GB) of the model editing methods per edit.}
        \label{fig:memory_cost}
    \end{minipage}
\vspace{-0.4cm}
\end{figure}

Figure~\ref{fig:time_cost} illustrates the average time cost per editing instance for each model editing method. Methods in the locate-then-edit category incur the highest time costs: PMET, the most costly, averages 21.2 seconds per edit across models, while ROME—more efficient than peers in this category—still averages 7.6 seconds per edit. Among all methods, MALMEN demonstrates the lowest time cost, with an average editing time of less than 1 second per edit. Memory-based methods and parameter-efficient fine-tuning approaches show similar time costs, ranging from 2.0 to 5.2 seconds per edit. AdaLoRA-L (averaging 3.8 seconds per edit) is faster than AdaLoRA (averaging 4.8 seconds per edit), primarily because it edits fewer parameters, resulting in reduced computational load. \revise{The layer identification step in AdaLoRA-L incurs a one-time offline cost of 52s, 117s, and 189s across models. Since this step is performed only once before editing and reused for all subsequent edits, its amortized overhead is negligible at approximately 0.1s per edit.} Figure \ref{fig:memory_cost} presents the average peak memory cost per editing instance for each method. PMET and AlphaEdit have the highest memory cost (12.7GB and 17.2GB on average across models), while MEMIT is the most memory-efficient at 10.1GB. The remaining methods have similar memory cost, averaging around 11.8 GB across the three models.

\begin{center} 
\vspace{-10pt}
\resizebox{\linewidth}{!}{
\begin{tabular}{!{\vrule width 1.5pt}p{1\columnwidth}}
     \makecell{{\large \faLightbulbO} \textbf{Answer to RQ2:}} The model editing methods generally have low time and memory overhead, allowing a single edit on a 3B-parameter model to complete within a few seconds on a 24 GB GPU.
\end{tabular}}
\vspace{-10pt}
\end{center}

\subsection{RQ3: How to improve the performance of AdaLoRA?}
\label{sec: rq3}

\subsubsection{Approach}

\begin{figure} [!t]
    \centering
    \includegraphics[width=0.95\linewidth]{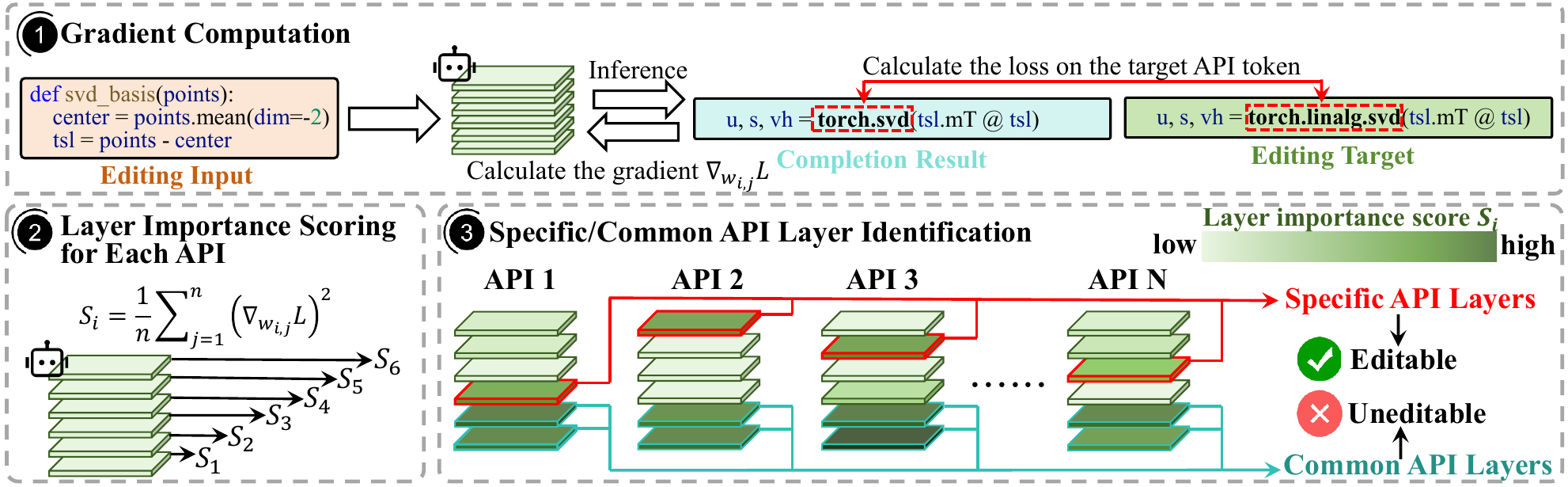}
    \captionsetup{skip=0pt}
    \caption{The identification process of Specific API Layers and Common API Layers.}
    \label{fig:adalora-l}
    \vspace{-0.65cm}
\end{figure}

To address RQ3, we aim to improve AdaLoRA, which achieves the highest Effectiveness, Generalization, and Portability among the methods explored in RQ1. Despite these strengths, AdaLoRA performs poorly in Specificity, likely because indiscriminate editing across all layers modifies parameters storing general knowledge, thereby disrupting information unrelated to the target API. 
A natural approach to enhance Specificity is to reduce the number of layers edited by AdaLoRA, thereby limiting the scale of parameter updates and avoiding unnecessary changes to general knowledge. However, selecting which layers to edit presents a critical challenge: failing to select layers relevant to the target API knowledge would degrade Effectiveness, Generalization, and Portability, while selecting layers that store general knowledge for editing would still compromise Specificity. To overcome this challenge, we propose AdaLoRA-L, an improved variant of AdaLoRA. It first precisely identifies layers storing general knowledge, then identifies API-specific layers for each target API from the remaining layers, and restricts editing operations exclusively to these API-specific layers. This design ensures that Specificity is enhanced without sacrificing performance in other dimensions. The overall layer selection process for AdaLoRA-L is illustrated in Figure~\ref{fig:adalora-l}.

\textbf{Gradient Computation:} For each editing instance, we have the LLM perform code completion on the editing input. We then calculate the loss based on the output—specifically, we calculate the loss on the target API token (i.e., the tokens corresponding to the up-to-date API in the editing target) while ignoring loss from other tokens.  Next, we backpropagate this loss to calculate the gradient for each editable parameter across all layers.
Consistent with neural network pruning methods~\cite{zhang2022platon,molchanov2019importance}, which employ gradients as a metric to quantify parameter importance with respect to specific samples, these computed gradients directly indicate a parameter’s relevance to the target API: a larger absolute gradient value signifies that even minor adjustments to the parameter would substantially impact the model’s capacity to output the correct up-to-date API token, meaning such parameters are critical for updating the current API.

\textbf{Layer Importance Scoring for Each API:} Using these parameter gradients, we further compute the layer importance score with the following formula: $S_{i} = \frac {1}{n}{\sum_{j = 1}^{n}\left ( {\nabla_{w_{i,j}} L} \right)^2}$, 
where  $S_{i}$ denotes the importance score of the \textit{i}-th layer, \textit{L} is the loss computed on the up-to-date API tokens, $w_{ij}$ represents the \textit{j}-th editable parameter in the \textit{i}-th layer, and \textit{n} is the total number of editable parameters in that layer. Intuitively, this score quantifies the average squared gradient magnitude of all editable parameters in the layer, reflecting the layer’s overall importance for enabling the model to predict the up-to-date API token.

For each editing instance, we calculate the importance scores of all layers. Then, for each API, we average the scores across all its editing instances to obtain API-specific layer importance scores, which indicate each layer’s relevance to the API. By analyzing the scores across all APIs, we identify:

\textbf{Common API Layers:} Layers that exhibit high importance across all APIs. These layers encode general knowledge critical to all APIs and are excluded from editing; 

\textbf{Specific API Layers:} Excluding Common API Layers, these are layers that demonstrate high importance only for the target API. These layers store the target API-specific knowledge and serve as the core targets for editing. 

Through this layer identification and selection process, AdaLoRA-L precisely isolates a set of specific API layers for each target API. These layers constitute the focused editing scope, ensuring parameter updates concentrate on API-specific knowledge while avoiding interference with general knowledge layers. This design enhances the Specificity of model editing and maximally preserves performance in Effectiveness, Generalization, and Portability.

\subsubsection{Results} 
\label{sec: rq3results}
\begin{table}[]
\footnotesize\tabcolsep=1.9pt
\caption{The performance comparison of of AdaLoRA-L, AdaLoRA, and REPLACEAPI.\vspace{-0.3cm}}
\label{tab:adaloral-performance}
\begin{tabular}{c|cccc|cccc|cccc|cccc}
\hline
\multirow{2}{*}{Editor} & \multicolumn{4}{c|}{Effectiveness} & \multicolumn{4}{c|}{Generalization} & \multicolumn{4}{c|}{Portability} & \multicolumn{4}{c}{Specificity} \\ 
\cline{2-17}
                        & AEM   & EM    & BU    & RL    & AEM   & EM    & BU    & RL    & AEM   & EM    & BU    & RL    & AEM   & EM    & BU    & RL    \\ 
\hline \rowcolor{lightgray}
% \multicolumn{17}{c}{DeepSeek-Coder} \\ 
% \cmidrule{1-17}
% Pre-editd               & 3.3   & 0.8   & 37.2  & 59.2  & 4.3   & 0.8   & 35.3  & 55.5  & 2.2   & 0.4   & 37.7  & 60.0  & 51.2  & 34.3  & 85.8  & 88.5  \\ 
% \cmidrule{1-17}
% AdaLoRA                 & 98.3  & 89.5  & 94.7  & 96.5  & \textbf{93.9} & \textbf{56.9} & 84.2  & 88.1  & 58.0  & 7.8   & 45.5  & 61.5  & 8.8   & 2.3   & 54.6  & 64.3  \\
% AdaLoRA-L               & \textbf{98.4} & \textbf{89.7} & \textbf{95.1} & \textbf{96.9} & 93.0  & 55.6  & \textbf{85.1} & \textbf{88.6} & \textbf{75.3} & \textbf{9.5} & \textbf{52.6} & \textbf{67.9} & \textbf{36.1} & \textbf{20.7} & \textbf{79.5} & \textbf{84.4} \\ 
% \cmidrule{1-17} \rowcolor{lightgray}
\multicolumn{17}{c}{Qwen2.5-Coder} \\ 
\hline
Pre-editd               & 2.7   & 0.7   & 40.4  & 63.4  & 5.1   & 1.0   & 36.9  & 58.2  & 2.6   & 0.7   & 39.8  & 62.9  & 99.3  & 93.7  & 97.6  & 97.2  \\ 
\hline
AdaLoRA                 & 98.1  & 60.4  & 81.0  & 87.6  & 89.5  & 39.3  & 69.9  & 77.7  & 71.4  & 11.7  & 49.7*  & 64.2*  & 49.2*   & 28.4*  & 47.9*  & 58.3*  \\
REPLACEAPI & 97.8              & 22.8*         & 67.4* & 79.2*   & 84.4*              & 17.9*         & 58.1* & 70.3*   & \textbf{97.2*}              & \textbf{22.8}         & \textbf{67.4*} & \textbf{78.6*}   & \textbf{99.2*}              & \textbf{93.5*}         & \textbf{97.4*} & \textbf{97.1*} \\
AdaLoRA-L               & \textbf{98.2} & \textbf{63.9} & \textbf{84.6} & \textbf{89.7} & \textbf{90.8} & \textbf{42.4} & \textbf{74.9} & \textbf{81.2} & 74.3 & 21.4 & 60.6 & 70.3 & 88.1 & 76.9 & 85.0 & 88.8 \\
\hline \rowcolor{lightgray}
\multicolumn{17}{c}{CodeGemma} \\ 
\hline
Pre-editd               & 1.0  & 0.4  & 37.6 & 60.8 & 4.5  & 1.6  & 29.4 & 49.4 & 0.8  & 0.3  & 36.2 & 59.7 & 99.5 & 98.3 & 99.2 & 96.8  \\ 
\hline
AdaLoRA                 & 99.0 & 91.7 & 95.3 & 96.7 & 78.7 & 39.4 & 65.1 & 71.3 & 74.2* & 13.6 & 49.2 & 61.9* & 51.2*  & 34.0*  & 46.5* & 56.4*  \\
REPLACEAPI & 98.2            & 21.7*        & 64.5* & 75.9*   & 66.2*            & 12.0*        & 44.9* & 57.4*   & \textbf{96.6*}            & \textbf{22.7}        & \textbf{63.3*} & \textbf{74.9}   & \textbf{98.0*}            & \textbf{96.3*}        & \textbf{97.5*} & \textbf{96.0*} \\
AdaLoRA-L               & \textbf{99.2} & \textbf{91.9} & \textbf{95.9} & \textbf{97.2} & \textbf{81.8} & \textbf{41.5} & \textbf{67.8} & \textbf{74.5} & 84.3 & 17.1 & 58.0 & 71.0 & 85.8 & 78.8 & 85.3 & 87.8 \\ 
\hline \rowcolor{lightgray}
\multicolumn{17}{c}{DeepSeek-Coder} \\ 
\hline
Pre-editd               & 3.3   & 0.8   & 37.2  & 59.2  & 4.3   & 0.8   & 35.3  & 55.5  & 2.2   & 0.4   & 37.7  & 60.0  & 86.4  & 78.7  & 85.8  & 88.5  \\ 
\hline
AdaLoRA                 & 98.3  & 89.5  & 94.7  & 96.5  & \textbf{93.9} & \textbf{56.9} & 84.2  & 88.1  & 58.0*  & 7.8   & 45.5  & 61.5  & 55.0*   & 37.5*   & 54.6*  & 64.3*  \\
REPLACEAPI & 98.0            & 13.3*        & 61.4* & 74.4*   & 78.8*            & 9.0*         & 50.7* & 64.1*   & \textbf{97.6*}            & \textbf{14.2*}        & \textbf{61.8*} & \textbf{75.1*}   & \textbf{86.3*}            & \textbf{78.5*}        & \textbf{85.9*} & \textbf{88.5} \\
AdaLoRA-L               & \textbf{98.4} & \textbf{89.7} & \textbf{95.1} & \textbf{96.9} & 93.0  & 55.6  & \textbf{85.1} & \textbf{88.6} & 75.3 & 10.3 & 52.6 & 67.0 & 80.0 & 69.1 & 79.5 & 84.4 \\ 
% \cmidrule{1-17} \rowcolor{lightgray}
\hline
\end{tabular}
\vspace{-0.6cm}
\end{table}

We set hyperparameters for AdaLoRA-L on the three models as follows: the number of frozen Common API Layers is set to 8, 8, and 4 for Qwen2.5-Coder, DeepSeek-Coder, and CodeGemma, respectively; the number of edited Specific API Layers is set to 4 for CodeGemma and 8 for the other two models. Table \ref{tab:adaloral-performance} presents the performance improvements of AdaLoRA-L compared to AdaLoRA. To evaluate the statistical significance of differences between AdaLoRA-L and baselines, we employ the Wilcoxon signed-rank test~\cite{Wilcoxon} with the Benjamini-Hochberg correction procedure~\cite{Haynes2013}. An asterisk (*) following the performance value indicates a statistically significant difference. 
In terms of Effectiveness, AdaLoRA-L achieves consistent performance gains across all three models.
In terms of Generalization, it
outperforms AdaLoRA on two models, with only a minor deficit observed on DeepSeek-Coder.
Notably, in both Portability and Specificity, AdaLoRA-L consistently surpasses AdaLoRA across all three LLMs. Specifically, Portability shows relative improvements of  4.1\%, 13.6\%, and 29.8\% on the three models in terms of AEM. Specificity exhibits substantial relative gains of 79.1\%, 67.6\%, and 45.5\% in AEM. Moreover, statistical tests confirm that these improvements are significant. 
This result clearly demonstrates that our proposed layer localization method not only substantially enhances the Specificity of AdaLoRA but also preserves its performance across the other three dimensions. 

\begin{center} 
\vspace{-10pt}
\resizebox{\linewidth}{!}{
\begin{tabular}{!{\vrule width 1.5pt}p{1\columnwidth}}
     \makecell{{\large \faLightbulbO} \textbf{Answer to RQ3:}} AdaLoRA-L, leveraging the localization of Specific API Layers and Common API Layers, accurately identifies the storage locations of knowledge for different APIs, enabling targeted editing of API knowledge. This approach not only effectively improves Specificity but also maximally retains performance in Effectiveness, Generalization, and Portability.
\end{tabular}}
\vspace{-10pt}
\end{center}

\subsection{RQ4: Is AdaLoRA-L more effective and efficient than REPLACEAPI?} 

We further compare AdaLoRA-L with \texttt{REPLACEAPI} \cite{wang2024llms}, a post-hoc approach that detects deprecated APIs in model outputs, removes the deprecated API and subsequent tokens, appends the corresponding up-to-date API, and feeds the modified prefix back to the LLM for regeneration. In terms of Effectiveness (AEM), AdaLoRA-L and \texttt{REPLACEAPI} achieve comparable performance. \revise{Although \texttt{REPLACEAPI} explicitly uses token-level replacement, it does not always reach 100\% AEM because the LLM may further modify the replaced API during subsequent regeneration.} For example, after substituting the deprecated \texttt{torch.eig} with \texttt{torch.linalg.eig}, the model may alter it to \texttt{torch.linalg.eigvalsh}. \revise{AdaLoRA-L achieves higher Generalization (AEM) than \texttt{REPLACEAPI}.} This is because a subset of generalization samples does not initially trigger deprecated APIs (as discussed in Section~\ref{sec: benchconstruction}); consequently, \texttt{REPLACEAPI} cannot apply replacements, whereas AdaLoRA-L, having been edited on similar contexts, directly generates the correct up-to-date APIs.
By contrast, \texttt{REPLACEAPI} achieves higher Portability (over 96\% AEM), while AdaLoRA-L attains 74.3\%–84.3\%. This gap arises because model editing is optimized on effectiveness data and may not fully generalize to substantially different contexts, i.e., portability data. \revise{At the same time, AdaLoRA-L yields consistently higher EM scores than REPLACEAPI in Effectiveness and
Generalization, with particularly large gaps in Effectiveness (41.1\%--76.4\%). This suggests that
AdaLoRA-L is more effective at generating complete and parameter-correct API usages, whereas
\texttt{REPLACEAPI} mainly guides the API-name replacement and still relies on the base LLM to complete the surrounding code.}
\revise{Thus, the results indicate a trade-off between post-hoc API-name correction and knowledge-level model editing, rather than showing that one approach dominates the other in all dimensions.}
While \texttt{REPLACEAPI} achieves higher Specificity by avoiding parameter updates, AdaLoRA-L nonetheless attains consistently high Specificity, with AEM exceeding 80\%. To further assess whether editing affects general coding ability, we evaluate the edited models on HumanEval~\cite{chen2021codex}. After AdaLoRA-L editing, pass@1 scores remain largely unchanged (46.3\%/31.7\%/29.3\% vs. 46.9\%/31.1\%/29.9\%), whereas AdaLoRA leads to a substantial degradation (36.6\%/23.7\%/26.2\%).
In terms of efficiency, \texttt{REPLACEAPI} incurs significantly higher inference cost, as it requires two rounds of generation. On EDAPIBench, its average inference time per sample is 3.31, 1.83, and 1.87 seconds across the three models, compared to 1.56, 0.89, and 0.87 seconds for AdaLoRA-L. \revise{This additional generation also increases token consumption, especially in long-context scenarios.}

\begin{center} 
\vspace{-10pt}
\resizebox{\linewidth}{!}{
\begin{tabular}{!{\vrule width 1.5pt}p{1\columnwidth}}
     \makecell{{\large \faLightbulbO} \textbf{Answer to RQ4:}}  \revise{AdaLoRA-L and REPLACEAPI show complementary strengths. AdaLoRA-L achieves stronger Generalization and parameter-correct API generation with lower inference cost, while REPLACEAPI provides high Portability and Specificity without modifying model parameters, at the cost of an additional regeneration step.}
\end{tabular}}
\vspace{-10pt}
\end{center}

\section{Discussion}

\begin{figure*}[t]
    \centering
    \begin{minipage}{0.28\textwidth}
        \centering
        \includegraphics[width=\textwidth]{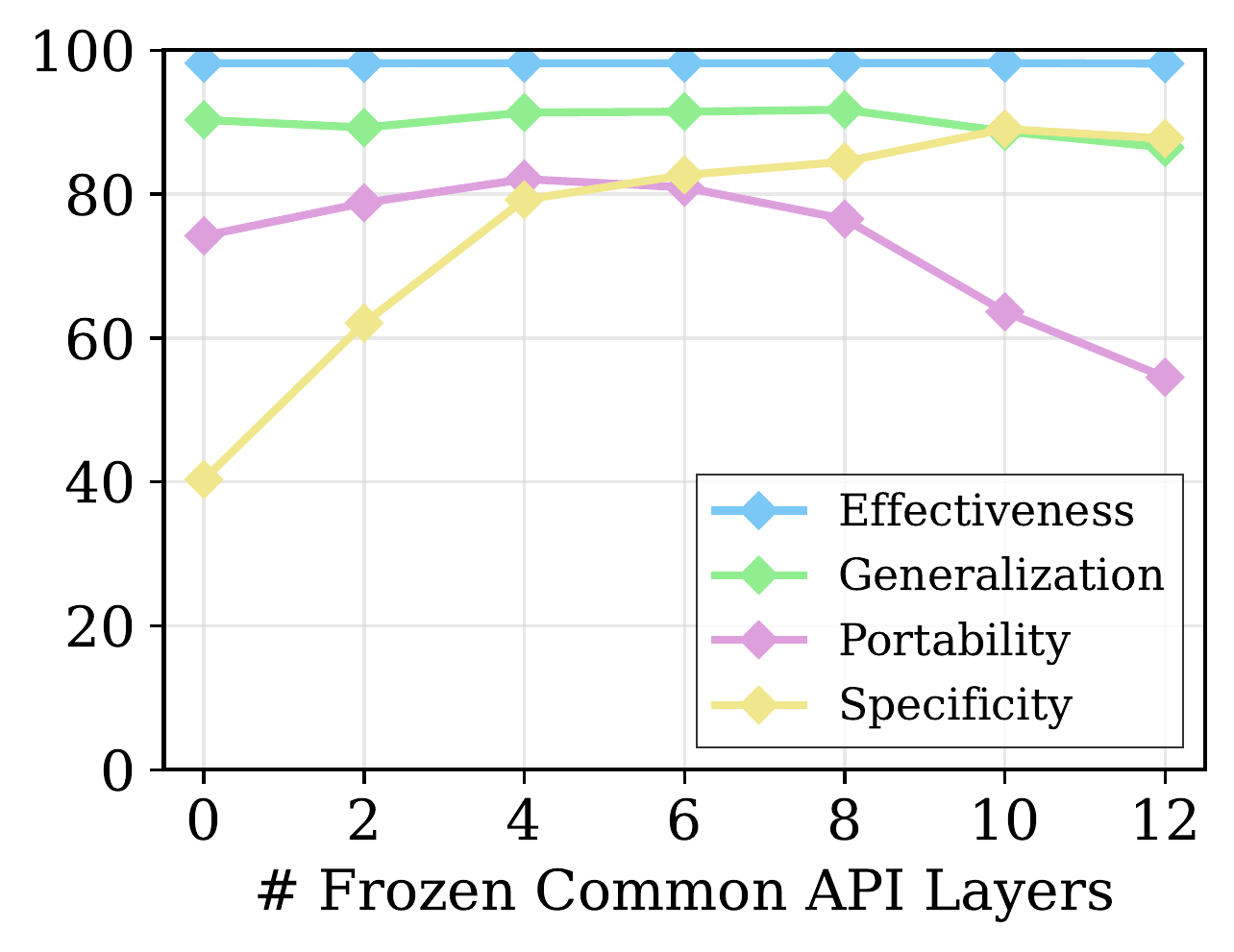}
        \captionsetup{skip=0pt}
        \caption*{Qwen2.5-Coder}
        \label{fig:qwencoder-common}
    \end{minipage}
    % 2. CodeGemma
    \begin{minipage}{0.28\textwidth}
        \centering
        \includegraphics[width=\textwidth]{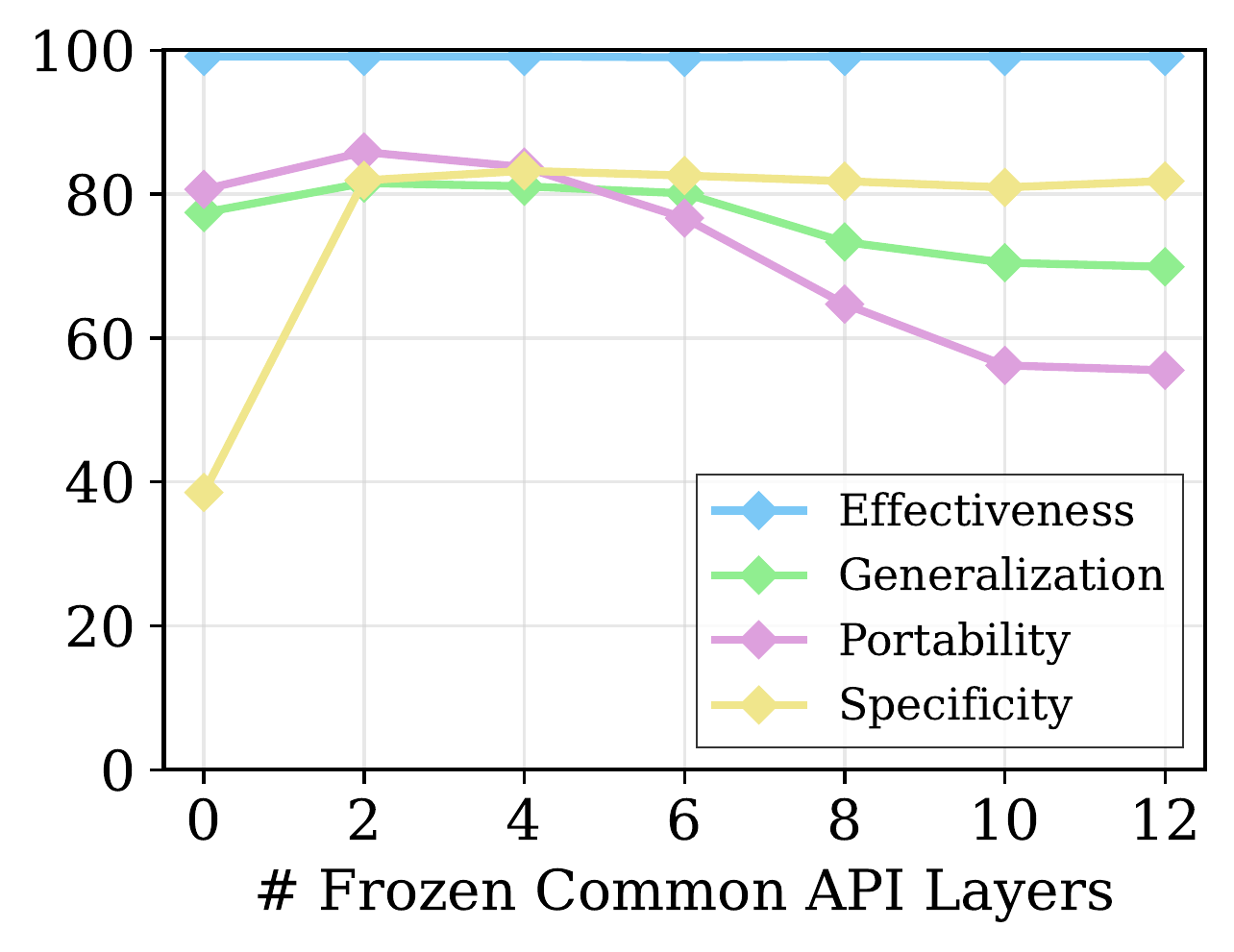}
        \captionsetup{skip=0pt}
        \caption*{CodeGemma}
        \label{fig:codegemma-common}
    \end{minipage}
    % 3. DeepSeek
    \begin{minipage}{0.28\textwidth}
        \centering
        \includegraphics[width=\textwidth]{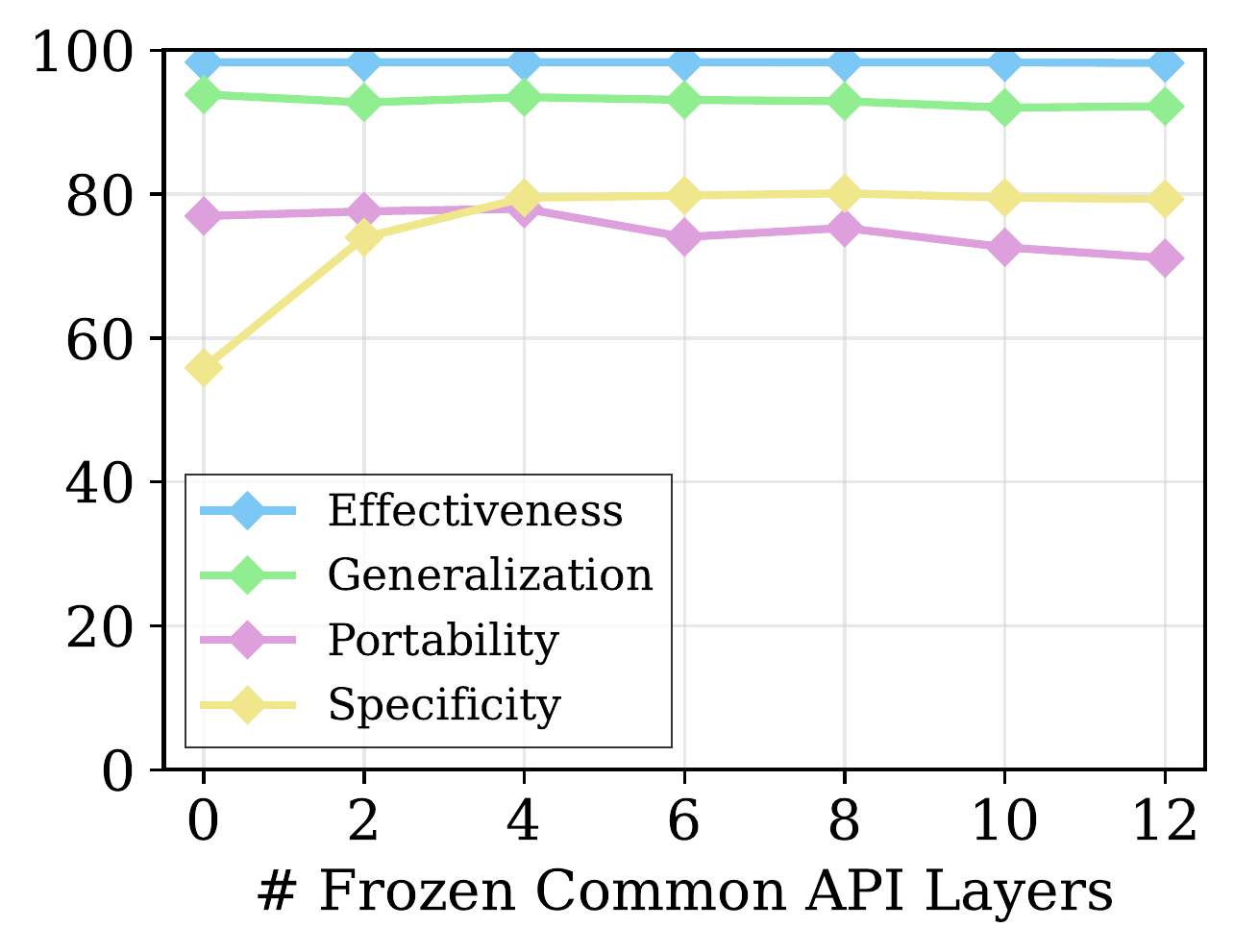}
        \captionsetup{skip=0pt}
        \caption*{DeepSeek-Coder}
        \label{fig:deepseek-common} 
    \end{minipage}
    
    \captionsetup{skip=5pt}
    \vspace{-0.2cm}
    \caption{The impact of the number of frozen Common API Layers on AdaLoRA-L.}
    \label{fig:common-layers}

    \begin{minipage}{0.28\textwidth}
        \centering
        \includegraphics[width=\textwidth]{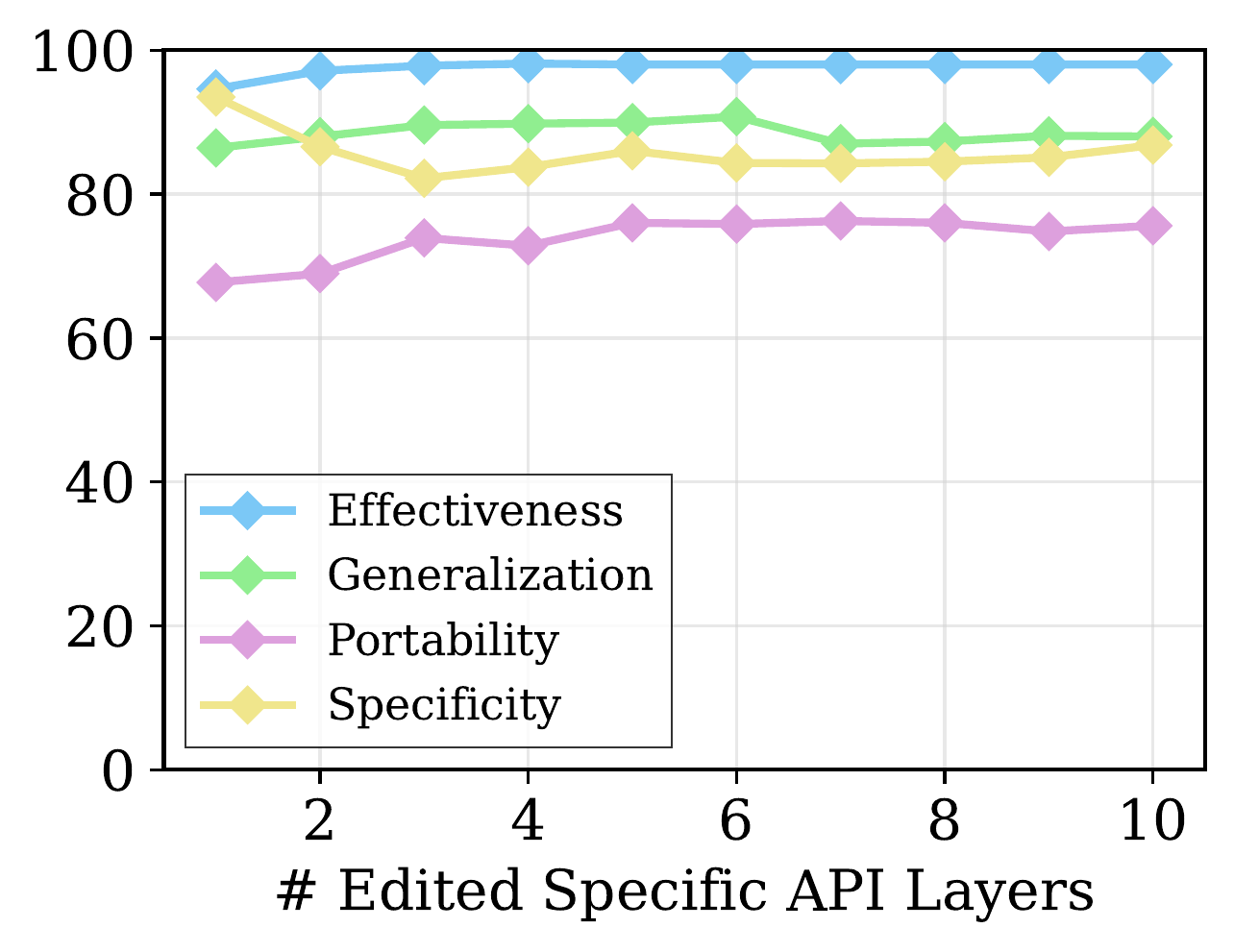}
        \captionsetup{skip=0pt}
        \caption*{Qwen2.5-Coder}
        \label{fig:qwencoder-specific}
    \end{minipage}
    % 2. CodeGemma
    \begin{minipage}{0.28\textwidth}
        \centering
        \includegraphics[width=\textwidth]{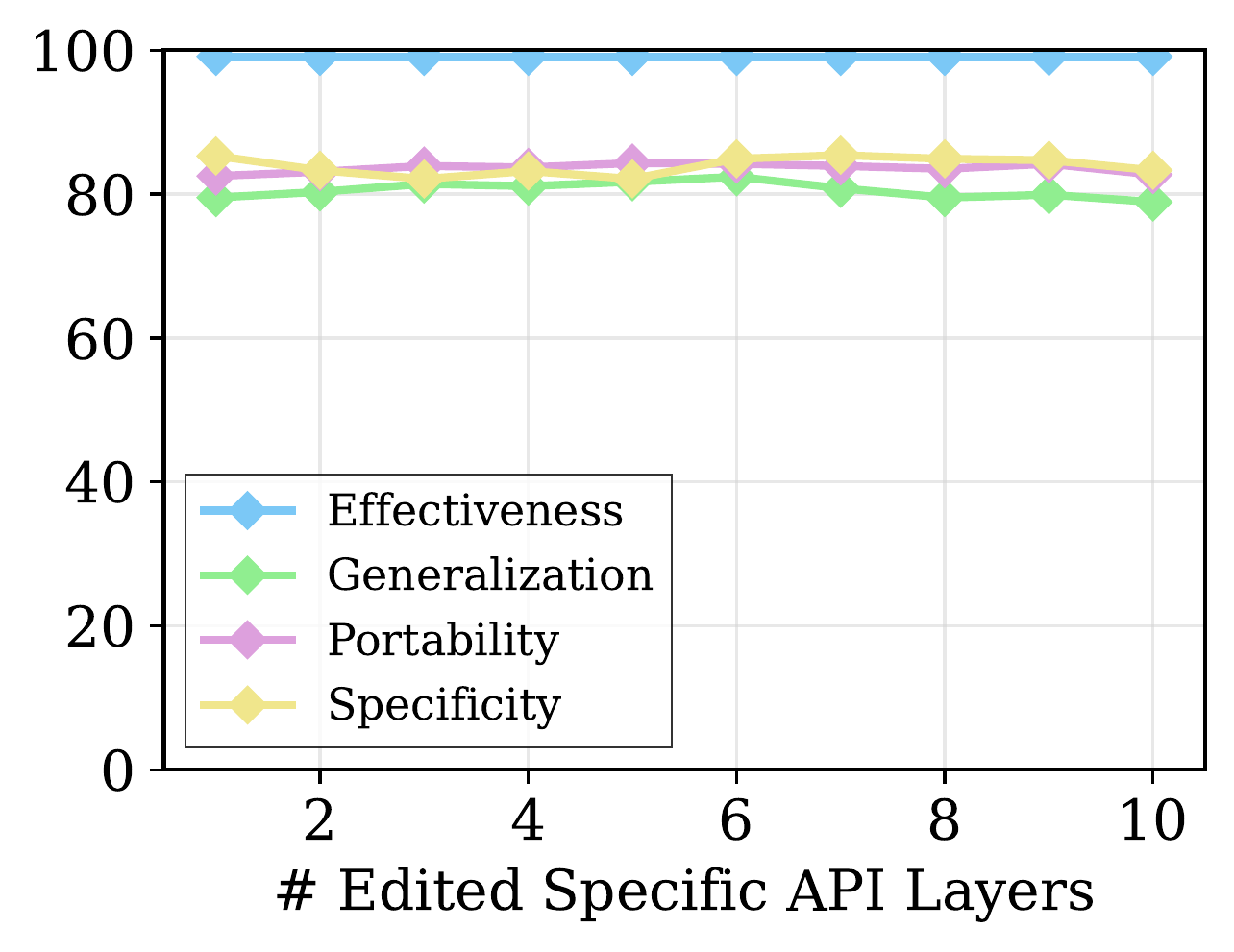}
        \captionsetup{skip=0pt}
        \caption*{CodeGemma}
        \label{fig:codegemma-specific}
    \end{minipage}
    % 3. DeepSeek
    \begin{minipage}{0.28\textwidth}
        \centering
        \includegraphics[width=\textwidth]{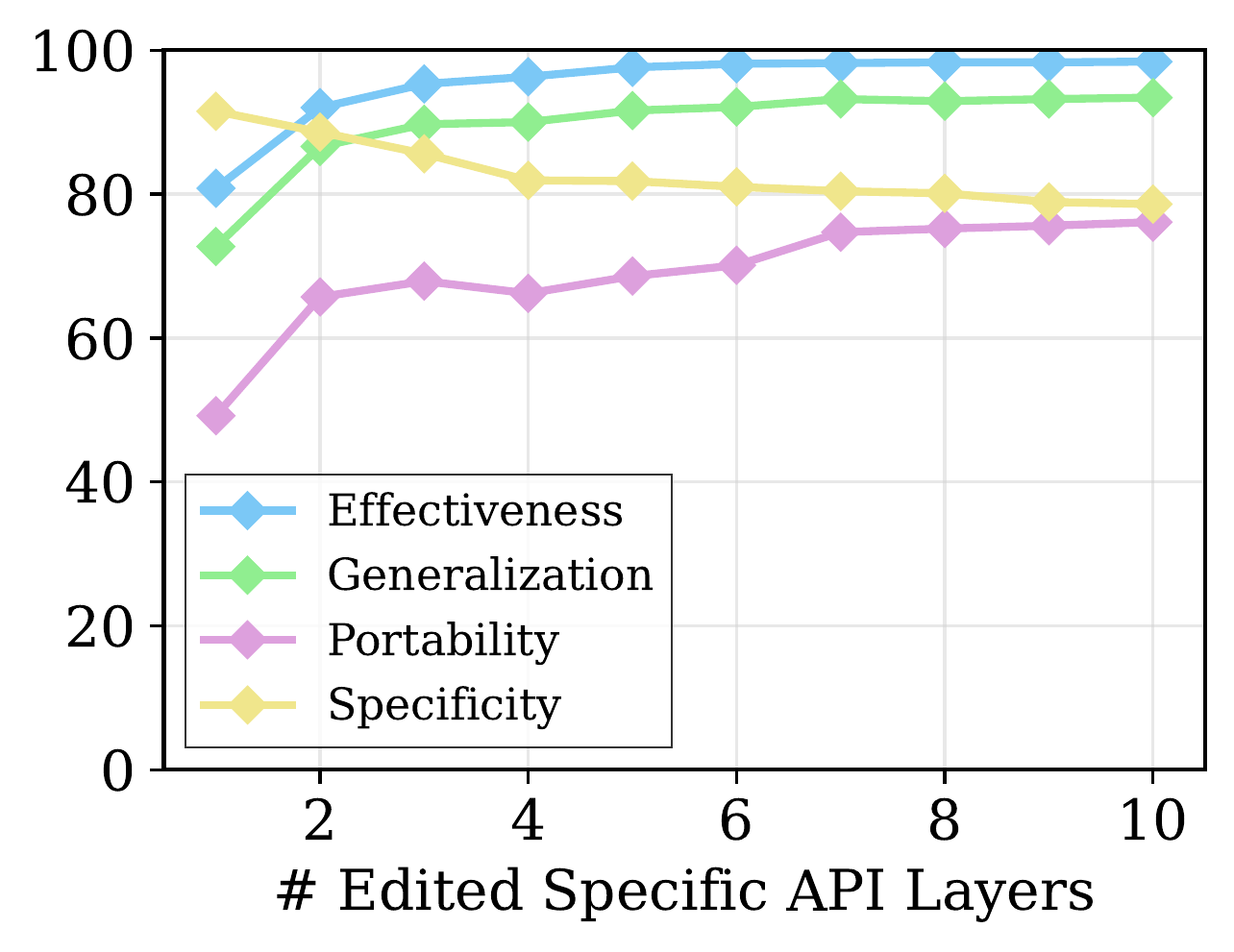}
        \captionsetup{skip=0pt}
        \caption*{DeepSeek-Coder}
        \label{fig:deepseek-specific} 
    \end{minipage}

    \captionsetup{skip=5pt}
    \vspace{-0.2cm}
    \caption{The impact of the number of edited Specific API Layers on AdaLoRA-L.}
    \label{fig:specific-layers}
    \vspace{-0.8cm}
\end{figure*}

\subsection{Impact of Hyperparameters on the Performance of AdaLoRA-L}
For RQ3, we set the number of frozen Common API Layers to 8 for DeepSeek-Coder and Qwen2.5-Coder. Given that CodeGemma has fewer total layers (18 compared to 24 for DeepSeek-Coder and 36 for Qwen2.5-Coder), we set this value to 4. Similarly, the number of edited Specific API Layers is set to 4 for CodeGemma and 8 for the other two models.
In this subsection’s experiment, we first fix the number of edited Specific API Layers at 4 for CodeGemma and 8 for the other two models, then vary the number of frozen Common API Layers within the range of 0–12. As shown in Figure \ref{fig:common-layers}, increasing the number of frozen Common API Layers enhances the model’s Specificity. Meanwhile, all models exhibit minimal changes in Effectiveness and Generalization—this is due to the fact that Generalization here targets edited inputs that only undergo syntactic rephrasing (the core of the input related to API editing remains unchanged), and both metrics depend on API-specific knowledge stored in Specific API Layers. The general knowledge in Common API Layers has low relevance to this kind of input scenario, so it has little impact on the two metrics. In contrast, Portability decreases as the number of frozen Common API Layers increases. Portability reflects the model’s ability to handle any input $x_p$, which represents a real-world input that is both semantically and syntactically different from the original editing input $x$ and is completed by the original LLM $f$ using the deprecated API $y_{d}$. Compared to the inputs for Generalization, such inputs require a greater amount of general knowledge to be processed effectively. Therefore, as more Common API Layers are frozen, the model’s Portability declines.

Subsequently, with the number of frozen Common API Layers fixed, we vary the number of edited Specific API Layers from 1 to 10. As shown in Figure~\ref{fig:specific-layers}, a consistent trend is observed across all three models (although it is less pronounced for CodeGemma): increasing the number of edited Specific API Layers—up to 8 for Qwen2.5-Coder and DeepSeek-Coder, and up to 4 for CodeGemma—leads to improvements in Effectiveness, Generalization, and Portability, while Specificity decreases. 
This behavior can be attributed to the expanded scope of parameter updates introduced by editing more layers. On the one hand, updating a larger number of parameters increases the likelihood of interfering with knowledge unrelated to the target API, resulting in reduced Specificity. On the other hand, involving more layers in the editing process enhances the model’s capacity to internalize the updated API knowledge, thereby improving editing success and facilitating better generalization to related inputs, which collectively boosts the three metrics. 
For CodeGemma, the gains from increasing the number of edited layers are comparatively limited. We attribute this to its relatively shallow architecture (18 layers, compared to 24 for DeepSeek-Coder and 36 for Qwen2.5-Coder), where semantic knowledge is more evenly distributed across layers. As a result, the difference in semantic capacity between single-layer and multi-layer editing is marginal, making it difficult for CodeGemma to benefit substantially from editing additional layers. Based on these observations and the need to balance performance across all evaluation dimensions, we adopt the final hyperparameter settings for the three models as described in Section~\ref{sec: rq3results} (RQ3).

\subsection{Impact of Model Scale on the Performance of AdaLoRA-L}

\begin{figure}[!t]
    \centering
    \begin{minipage}{0.32\textwidth}
        \centering
        \includegraphics[width=\textwidth]{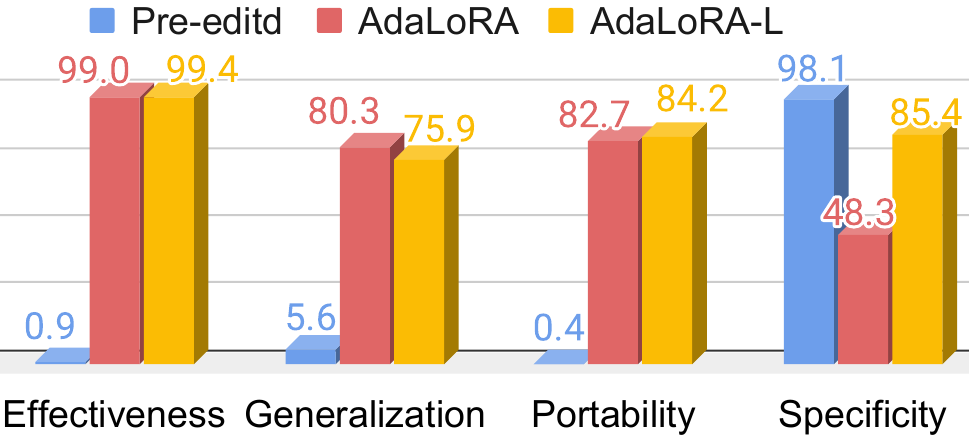}
        \captionsetup{skip=0pt}
        \caption*{Qwen2.5-Coder (1.5B)}
        \label{fig:1.5B-adaloral}
    \end{minipage}
    \begin{minipage}{0.32\textwidth}
        \centering
        \includegraphics[width=\textwidth]{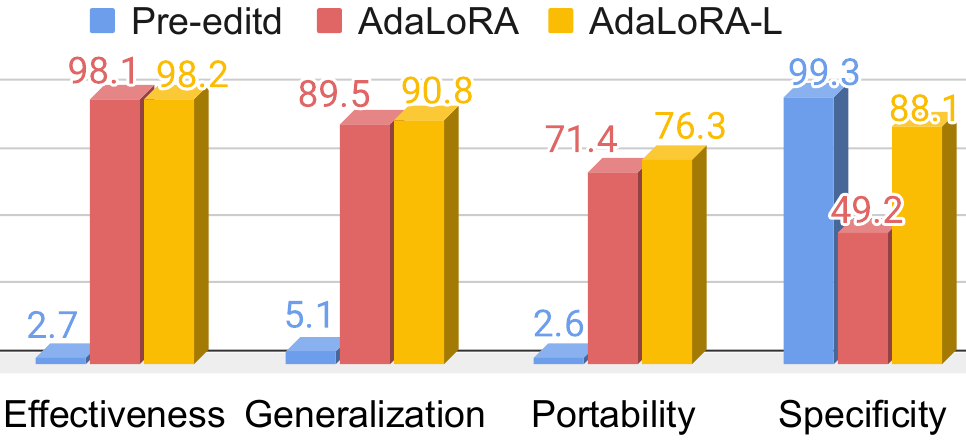}
        \captionsetup{skip=0pt}
        \caption*{Qwen2.5-Coder (3B)}
        \label{fig:3B-adaloral}
    \end{minipage}
    \begin{minipage}{0.32\textwidth}
        \centering
        \includegraphics[width=\textwidth]{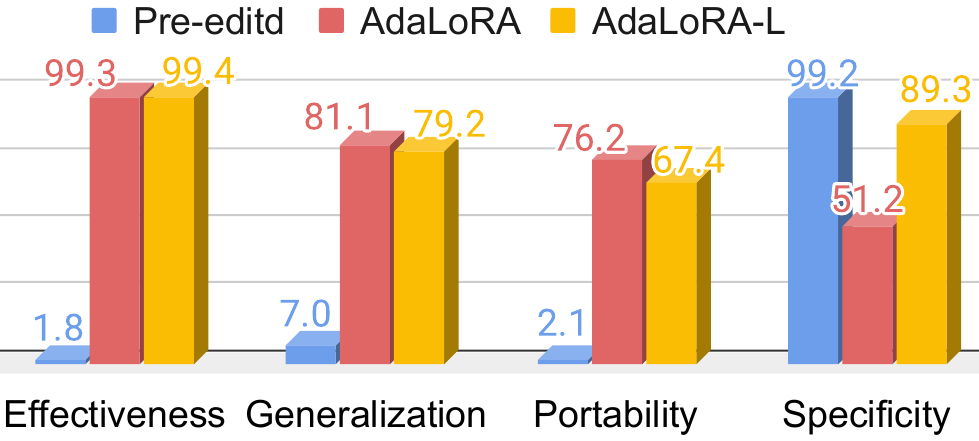}
        \captionsetup{skip=0pt}
        \caption*{Qwen2.5-Coder (7B)}
        \label{fig:7B-adaloral} 
    \end{minipage}
    \captionsetup{skip=0pt}
    \caption{The performance of AdaLoRA-L across different scales of Qwen2.5-Coder.}
    \label{fig:different-size-llm}
    \vspace{-0.45cm}
\end{figure}

This section investigates whether the LLM scale influences AdaLoRA-L’s performance. We select the Qwen2.5-Coder series due to its more diverse range of model sizes compared to CodeGemma and DeepSeek-Coder, enabling us to test AdaLoRA-L and AdaLoRA across three distinct scales: 1.5B, 3B, and 7B parameters. The results in Figure ~\ref{fig:different-size-llm}  demonstrate that AdaLoRA-L consistently achieves stable improvements in Specificity compared to AdaLoRA across all model scales, while maintaining robust performance across the remaining three dimensions regardless of model size.

\subsection{Impact of Different Libraries on Model Editing Performance}

\begin{figure}
    \centering

    \begin{minipage}[t]{0.49\textwidth}
        \centering
        \includegraphics[width=\textwidth]{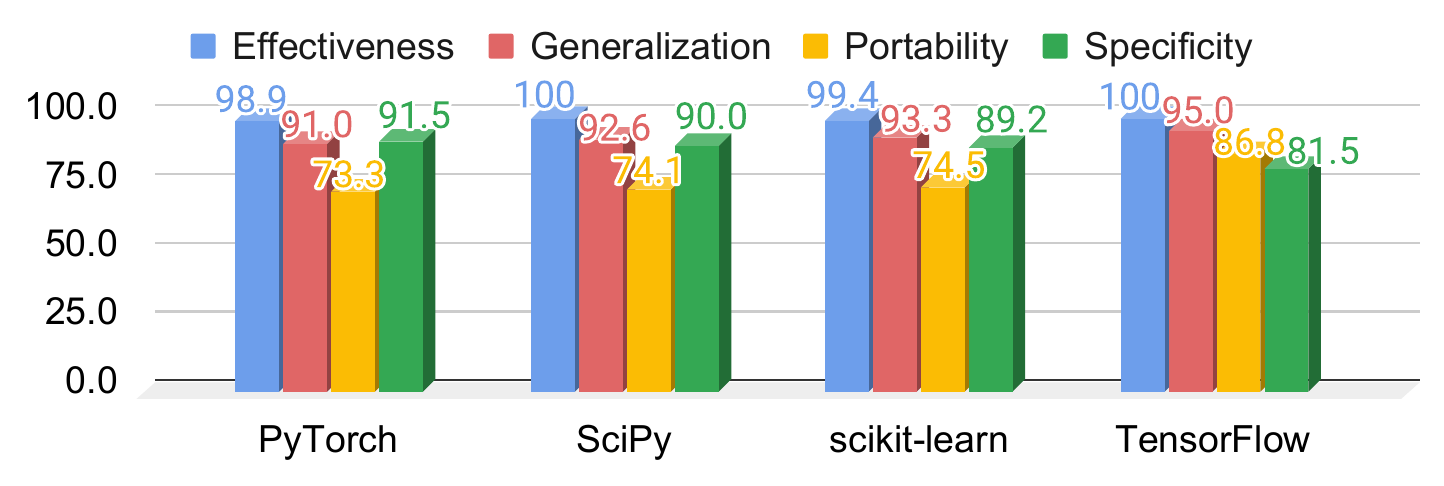}
        \captionsetup{skip=0pt}
        \caption{The performance of AdaLoRA-L across different libraries on Qwen2.5-Coder.}
        \label{fig:qwencoder-lib}
    \end{minipage}
    \begin{minipage}[t]{0.49\textwidth}
        \centering
        \includegraphics[width=\textwidth]{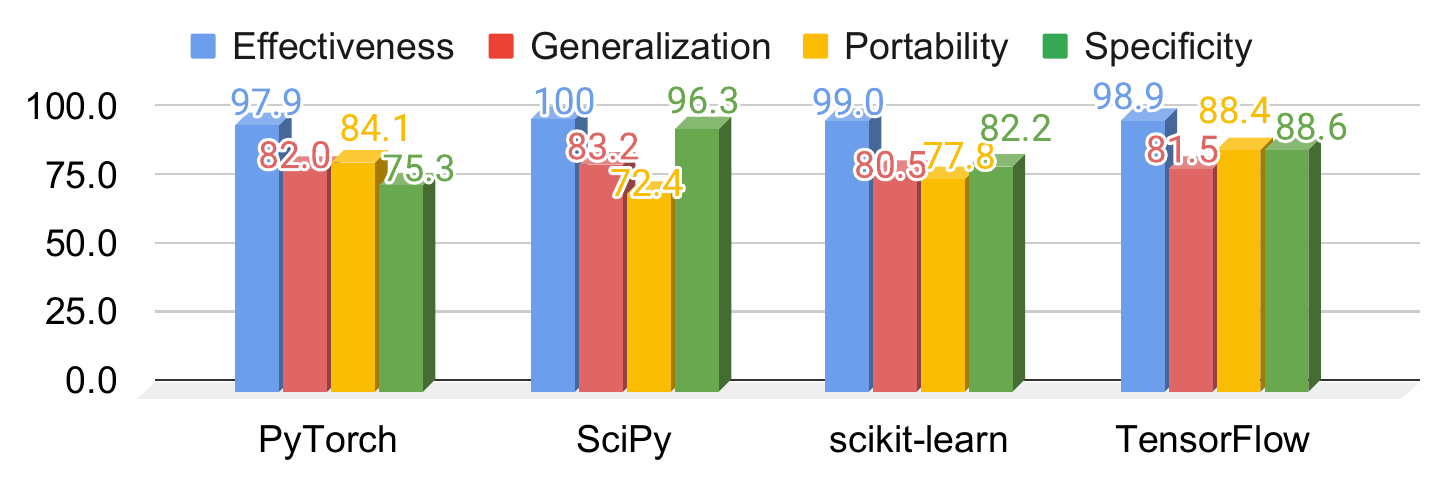}
        \captionsetup{skip=0pt}
        \caption{The performance of AdaLoRA-L across different libraries on CodeGemma.}
        \label{fig:codegemma-lib}
    \end{minipage}

    \begin{minipage}[t]{0.49\textwidth}
        \centering
        \includegraphics[width=\textwidth]{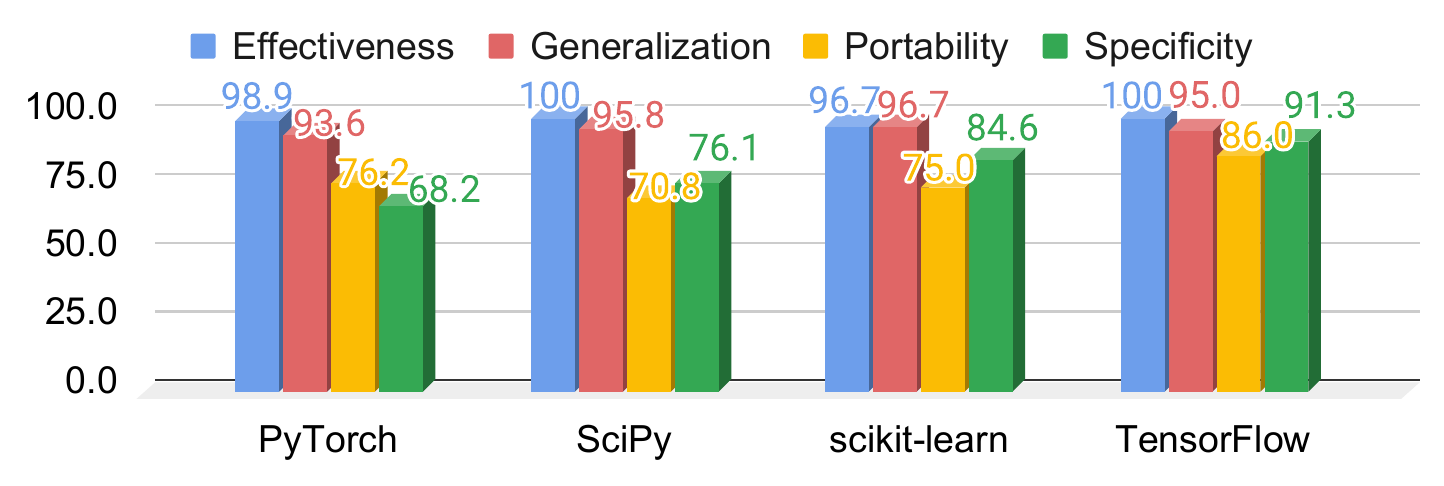}
        \captionsetup{skip=0pt}        
        \caption{The performance of AdaLoRA-L across different libraries on DeepSeek-Coder.}
        \label{fig:deepseek-lib} 
    \end{minipage}
    \begin{minipage}[t]{0.49\textwidth}
        \centering
        \includegraphics[width=\textwidth]{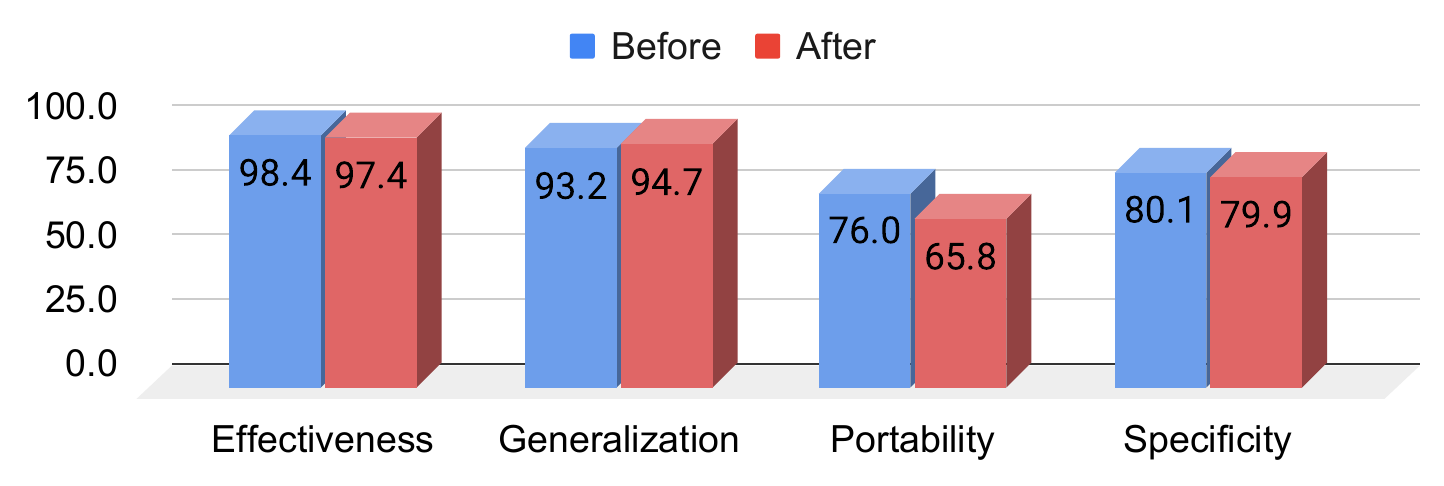}
        \captionsetup{skip=0pt}
        \caption{The impact of API deprecation timing (Before vs. After LLM training data cutoff) on AdaLoRA-L.}
        \label{fig:deprecated_timming}
    \end{minipage}
\vspace{-0.6cm}
\end{figure}

This section analyzes the editing performance of AdaLoRA-L on APIs from different libraries. To ensure statistical significance, we focus on the four most sample-rich libraries: PyTorch, SciPy, scikit-learn, and TensorFlow. Figures  \ref{fig:qwencoder-lib}, \ref{fig:codegemma-lib}, and \ref{fig:deepseek-lib} present the API Exact Match (AEM) results across the four evaluation dimensions for the three models, respectively.
Our observations reveal that Effectiveness and Generalization show little variation across libraries. In contrast, Portability and Specificity exhibit more noticeable differences across libraries. For example, on DeepSeek-Coder and CodeGemma, PyTorch achieves notably lower Specificity scores (68.2 and 75.3, respectively) compared to TensorFlow (91.3 and 88.6). Across all three models, SciPy consistently exhibits the lowest Portability, averaging 72.4, whereas TensorFlow achieves the highest average Portability of 87.1.  
Overall, while no strong or consistent patterns emerge across libraries, AdaLoRA-L demonstrates robust performance across all four evaluation dimensions for these libraries, confirming its capability to effectively edit deprecated API knowledge regardless of the library.

\subsection{Impact of API Deprecation Timing on Model Editing Performance}

This section investigates whether the timing of API deprecation affects the performance of deprecated API knowledge editing. We first analyze the official deprecation dates of each deprecated API in \texttt{EDAPIBench}, categorizing them as deprecated either before or after each model's training data cutoff. Statistics show all APIs are deprecated before CodeGemma and Qwen2.5-Coder's cutoffs. For DeepSeek-Coder, 6 APIs (38 editing instances) are deprecated post-cutoff, while 70 APIs (1,033 instances) are deprecated pre-cutoff. Thus, Figure~\ref{fig:deprecated_timming} presents AdaLoRA-L results on DeepSeek-Coder across these two groups in terms of AEM.  The results show that the deprecation timing relative to DeepSeek-Coder’s cutoff has minimal impact on Effectiveness, Generalization, or Specificity. However, Portability is slightly lower for APIs deprecated post-cutoff. A plausible explanation is as follows: during DeepSeek-Coder’s pre-training, the model only encountered non-deprecated usages of these 6 post-cutoff APIs, without exposure to their deprecation status or any updated alternatives. In contrast, for the 70 pre-cutoff APIs, the training data likely included both deprecated and up-to-date usages. This discrepancy causes DeepSeek-Coder to form a strongly entrenched understanding of the 6 post-cutoff APIs as valid, non-deprecated entities. Consequently, when updating the deprecated API knowledge for these 6 APIs, the newly injected information must counteract the model’s rigid pre-existing understanding of ``valid usage''—rendering this new knowledge less robust and ultimately resulting in slightly lower Portability.

% \begin{figure}
%     \centering
%     \includegraphics[width=\textwidth]{figures/api_deprecated_date_impact.pdf}
%     \caption{Impact of API Deprecation Timing (Before vs. After LLM Training Data) on Editing Performance.}
%     \label{fig:deprecated_timming}
% \end{figure}

\subsection{When to Use AdaLoRA-L and REPLACEAPI}

\revise{The comparison with REPLACEAPI highlights a practical trade-off between model editing and post-hoc API replacement. REPLACEAPI has several attractive properties: it is model-agnostic, does not require access to model parameters, and can be deployed whenever a deprecated-to-up-to-date API mapping is available. Since it does not modify the model, it naturally preserves unrelated model behavior, which explains its strong Specificity results. Therefore, REPLACEAPI is a suitable choice when developers mainly need to prevent deprecated API names from appearing in generated code, when the target model is closed-source or cannot be edited, or when deployment simplicity is more important than reducing per-query inference cost. AdaLoRA-L is more suitable when the goal is to update the model’s internal API knowledge rather than only rewrite the generated API name. This distinction becomes important when API evolution changes function signatures, parameter names, return values, or expected usage patterns. In these cases, replacing the API name alone may still leave the LLM responsible for generating the correct surrounding code from outdated knowledge. AdaLoRA-L is also preferable in interactive or high-throughput settings, such as IDE code completion, where repeated detection and regeneration can increase latency and token consumption. Thus, the two approaches should be viewed as serving different deployment needs. REPLACEAPI offers a non-invasive solution for API-name-level correction, especially for closed-source models or low-frequency use cases. AdaLoRA-L is better suited for open-source models and production settings that require parameter-correct API usage and lower inference overhead.}

\section{Threats of Validity}

(1) Even with fixed temperature and greedy decoding, LLM outputs vary across runs. This causes occasional non-zero pre-edit AEM scores and Specificity AEM rarely reaching 100\%. To reduce these effects, each experiment is repeated five times, and the median is reported.
(2) Our study focuses on single-line API completion—checking whether the edited LLM generates the correct up-to-date API—rather than multi-line code or end-to-end functional correctness. This is because (a) multi-line correctness requires task-specific test cases, difficult to standardize across 3,000+ instances, and (b) prompts often lack enough context for the LLM to infer full multi-line logic, making evaluation prone to bias. Accordingly, following established practices in code completion studies~\cite{yu2024fight, zhang2023repocoder}, we focus on single-line evaluation.
(3) For AdaLoRA-L, Effectiveness and Generalization EM are high, but Portability EM is comparatively lower, despite relatively strong BLEU and ROUGE-L scores. This may be partly because the LLM generates syntactically valid API calls with arguments that do not match the ground truth—a limitation also observed in existing single-line code-completion methods, where EM rarely exceeds 0.3 for LLMs with up to 3B parameters~\cite{Zhang_Li_Li_Xia_Yang_Luo_Wang_Chen_Liu_Qu_Yang_2025}. Another contributing factor is that model editing is performed only on Effectiveness data, leaving the model undertrained on Portability scenarios. Enhancing Portability EM is a promising direction for future work, for example by providing updated API descriptions or documentation during editing to improve the model’s ability to generalize API usage.
\revise{(4) Our benchmark currently relies on deprecated-to-up-to-date API mappings from Wang et al.~\cite{wang2024llms}. Extending it to additional new APIs would require manually collecting new mappings.}

\section{Related Works}

\textbf{API Evolution.} 
Prior studies~\cite{sawant2018features,sawant2018understanding,mirian2019web,sawant2019react} have examined the motivations behind API deprecation and how developers respond to the changes. Common reasons for deprecating APIs include improving code readability, reducing redundancy, addressing poor coding practices, and fixing bugs. Deprecated APIs can impact hundreds of dependent projects~\cite{robbes2012developers}, particularly as developers struggle to keep pace with fast-evolving software ecosystems~\cite{linares2013api,wang2020empirical}. For instance, McDonnell et al.~\cite{mcdonnell2013empirical} found that only 22\% of deprecated API usages are eventually migrated to replacement APIs. Hora et al.~\cite{hora2015developers} observed that while developers invest considerable effort in identifying and adopting alternatives, most projects fail to take timely action to address deprecations. Such delays lead to the silent accumulation of technical debt, complicating future migrations as multiple API changes must be resolved simultaneously~\cite{sawant2016reaction}. Against this backdrop, deprecated APIs pose a critical problem for LLMs—a challenge amplified by their growing integration into software development workflows. 
Wang et al.~\cite{wang2024llms} found that 37.4\% of GPT-3.5’s API predictions are deprecated, motivating our work to update LLMs’ API knowledge via model editing to keep pace with evolving APIs.

\begin{figure}[!t]
    \centering
    \includegraphics[width=0.98\linewidth]{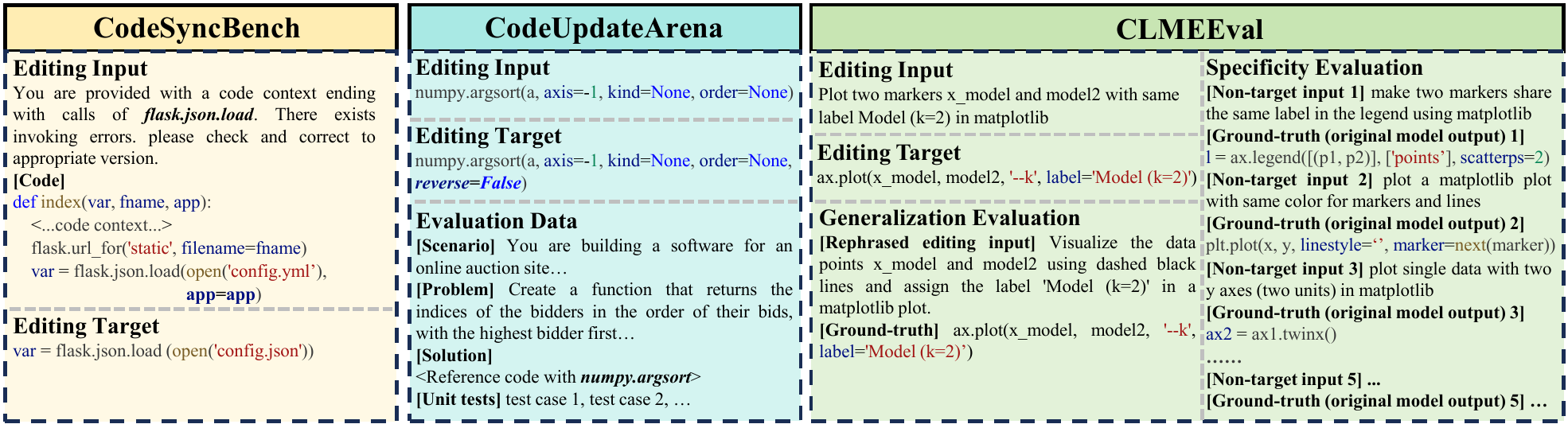}
    \captionsetup{skip=0pt}
    \caption{The three benchmarks similar to our \texttt{EDAPIBench}.}
    \label{fig:benchmark_comparation}
    \vspace{-0.8cm}
\end{figure}

\textbf{LLM API Knowledge Updating.} 
In the broader domain of LLM API knowledge updating, three closely related studies are Wang et al.’s \texttt{CodeSyncBench}~\cite{wang2025codesync}, Liu et al.’s \texttt{CodeUpdateArena}~\cite{liu2024codeupdatearena}, and Chhetri et al.'s empirical study~\cite{chhetri2025understanding}. The first two focus on evaluating LLMs’ ability to adapt to API parameter changes (e.g., addition or removal of parameters) and generate code with correct parameter usage. \texttt{CodeSyncBench} focuses on API parameter correction tasks, such as fixing code like 
\texttt{var= flask.json.load(open(`config.yml'), app=app)}
to the correct form \texttt{var = flask.json.load(open(`config.json'))} 
(Figure \ref{fig:benchmark_comparation}) 
to test whether LLMs can identify and fix parameter mismatches. 
\texttt{CodeUpdateArena} employs LLM-synthesized code generation tasks that do not fully reflect real-world scenarios and often rely on supposedly updated API parameters. For example, one task instructs the LLM to generate code using \texttt{numpy.argsort} with a newly introduced \texttt{reverse} parameter (Figure~\ref{fig:benchmark_comparation}). However, \texttt{numpy.argsort} does not actually support a \texttt{reverse} parameter, indicating that such tasks are based on synthetic and unrealistic API specifications.

Regarding methods for updating API parameter knowledge, Liu et al.~\cite{liu2024codeupdatearena} relied solely on LoRA, which yielded limited Effectiveness: the maximum improvement in Pass@1 reached only 6\%. \revise{Wu et al.'s \texttt{ReCode}~\cite{wu2026recode} investigated rule-based reinforcement learning (RL) for API knowledge updating and reported substantial Pass@1 improvements on \texttt{CodeUpdateArena}.} Wang et al.~\cite{wang2025codesync}, meanwhile, employed SFT-LoRA~\cite{peng2023instruction} alongside three RL-based approaches: DPO~\cite{rafailov2023direct}, SimPO~\cite{meng2024simpo}, and ORPO~\cite{hong2024orpo}. While these RL methods also use LoRA for fine-tuning, they differ from standard LoRA in their fine-tuning signals: RL relies on reward-driven feedback (e.g., DPO’s preference rankings, ORPO’s pairwise comparisons) rather than direct supervised signals for targeted knowledge updates. 
Wang et al.~\cite{wang2025codesync} found that DPO achieved the best performance, improving CodeBLEU scores from 32.68 (original model) to 44.95, though the absolute improvement remained modest. They also noted that RL methods required two to three times more training time than standard LoRA and involved additional reward engineering. These factors conflict with our focus on lightweight, efficient editing of deprecated API knowledge, where simplicity and speed are prioritized. Consequently, our study concentrates exclusively on LoRA-based techniques and does not incorporate RL-based approaches. 

Another key distinction between these benchmarks and our \texttt{EDAPIBench} lies in the evaluation of Specificity. Both \texttt{CodeSyncBench} and \texttt{CodeUpdateArena} assess Specificity indirectly by tracking changes in Pass@1 scores on the HumanEval benchmark after fine-tuning. However, we argue that a more rigorous evaluation—aligned with \texttt{EDAPIBench}’s design—should directly verify whether updated LLMs can correctly generate parameters for other non-updated APIs within \texttt{CodeSyncBench} and \texttt{CodeUpdateArena}. This direct assessment ensures that the knowledge update process does not inadvertently degrade the model’s existing knowledge of unaffected APIs, providing a more precise and targeted measure than indirect metrics such as HumanEval Pass@1. 

Meanwhile, concurrent work by Chhetri et al.~\cite{chhetri2025understanding} also focuses on addressing API deprecation using five editing methods (all covered by the ten methods we investigate). They select 797 HumanEval and MBPP tasks with API calls and create synthetic deprecation scenarios—for example, replacing \texttt{math.sqrt()} (treated as deprecated) with \texttt{math.square\_root()} (treated as an up-to-date API). Notably, these scenarios are also unrealistic, as \texttt{math.square\_root()} does not exist in reality.

\textbf{Model Editing. } 
Model editing is an active NLP research area, with many techniques and benchmarks developed to refine general-purpose LLMs. For example, $\text{WikiData}_\text{recent}$~\cite{cohen2024evaluating}, ZsRE~\cite{levy2017zero}, and KnowEdit~\cite{zhang2024comprehensive} are widely used benchmarks for factual knowledge editing. However, a key limitation of these datasets is that they do not verify whether models actually produce incorrect or hallucinated outputs before model editing. As a result, using these benchmarks to evaluate post-editing performance makes it difficult to accurately assess the effectiveness of different knowledge editing techniques in correcting hallucinations. In contrast, our \texttt{EDAPIBench} triggers edits only when models actually generate deprecated APIs, providing a more realistic and targeted evaluation scenario. 
The application of model editing to software engineering tasks remains relatively limited.  Gu et al.~\cite{gu2024neuronpatchingsemanticbasedneuronlevel} proposed MENT, which repaired next-token errors in code generation by patching specific neurons in LLMs. Liu et al.~\cite{liu2025improving} developed CREME to enhance LLMs’ robustness against prompt perturbations in code generation (e.g., typos) via targeted parameter updates in robustness-sensitive layers. Nevertheless, these approaches focus on code generation tasks where there is no fixed correct output. This differs from our focus on factual knowledge updating, and thus these methods are not included in our evaluation. A more closely related work is Li et al.’s \texttt{CLMEEval}~\cite{li2024model}—a model editing benchmark derived from \texttt{CoNaLa}~\cite{yin2018learning} and \texttt{CodeSearchNet}~\cite{husain2019codesearchnet}. However, the task setup in CLMEEval is somewhat detached from practical scenarios. For example (as illustrated in Figure \ref{fig:benchmark_comparation}), given a simple plotting description (e.g., “\textit{Plot two markers x\_model and model2 with same label Model (k=2) in matplotlib
}”) as the editing input, the benchmark directly sets the target output to \texttt{ax.plot(x\_model, model2, `----k', label=`Model (k=2)')}
without first verifying whether the LLM already generates the correct code from the prompt prior to editing. This diverges from a realistic editing paradigm, where edits are applied only when the model produces errors. Furthermore, CLMEEval relies on exact-match evaluation, which is limiting for code generation tasks since LLMs may generate functionally correct code that differs syntactically from the target. The benchmark also evaluates generalization solely through simple input paraphrases (e.g., textual rewrites of the original task), with limited diversity in paraphrase types. In contrast, our benchmark triggers editing only when LLMs output deprecated APIs. We introduce an “API Exact Match” metric to more precisely measure editing Effectiveness. To assess Generalization, we utilize GPT-4.1-generated code rewrites and incorporate a Portability dimension that evaluates whether the edited model can correctly complete the updated API calls across different editing instances, where the original model also completes the prompt with the same deprecated API.

\section{Conclusion and Future Works}

This study presents \texttt{EDAPIBench}, a \revise{mostly} automated benchmark for editing deprecated API knowledge in LLMs, covering over 70 deprecated APIs from 8 Python libraries with 900+ instances per model. Unlike most existing benchmarks in NLP and software engineering, it only includes cases where LLMs initially generate deprecated APIs, ensuring meaningful needs-editing evaluations. Thus, \texttt{EDAPIBench} provides a standardized, rigorous platform for model editing research, benefiting both the software engineering and broader NLP communities.
We evaluate 10 editing methods, finding AdaLoRA most effective but limited in Specificity, motivating our AdaLoRA-L variant. Overall, AdaLoRA-L offers a lightweight approach for updating deprecated API knowledge, enabling LLMs to generate more reliable and up-to-date code without disrupting other API behaviors. Future work will expand \texttt{EDAPIBench} to cover more deprecated APIs across additional programming languages. 

\section{Data Availability}
Our \texttt{EDAPIBench}, the code for its construction, and model editing source code are available in~\cite{opensource}. 

\section{Acknowledgments}

This research was supported by National Natural Science Foundation of China under Grant No. 62502440 and Zhejiang Provincial Natural Science Foundation of China under Grant No. LQN26F020003.

\bibliographystyle{ACM-Reference-Format}
\bibliography{main/references}
\end{document}